\documentclass[12pt]{article}
\usepackage{graphicx}

\begin{document}

\title{Application of Abel-Plana formula
for collapse and revival of Rabi oscillations
in Jaynes-Cummings model}

\author{Hiroo Azuma\thanks{On leave from
Institute of Computational Fluid Dynamics,
1-16-5 Haramachi, Meguro-ku, Tokyo 152-0011, Japan.}
\\
{\small Information and Mathematical Science Laboratory Inc.,}\\
{\small Meikei Bldg., 1-5-21 Ohtsuka, Bunkyo-ku,
Tokyo 112-0012, Japan}\\
{\small E-mail: hiroo.azuma@m3.dion.ne.jp}
}

\date{20 July 2010}

\maketitle

\begin{abstract}
In this paper, we give an analytical treatment
to study the behavior of the collapse and the revival of the Rabi oscillations
in the Jaynes-Cummings model (JCM).
The JCM is an exactly soluble quantum mechanical model,
which describes the interaction between a two-level atom
and a single cavity mode of the electromagnetic field.
If we prepare the atom in the ground state and the cavity mode in a coherent state initially,
the JCM causes the collapse and the revival of the Rabi oscillations many times
in a complicated pattern in its time-evolution.
In this phenomenon, the atomic population inversion is described with an intractable infinite series.
(When the electromagnetic field is resonant with the atom,
the $n$th term of this infinite series is given by a trigonometric function for $\sqrt{n}t$,
where $t$ is a variable of the time.)
According to Klimov and Chumakov's method,
using the Abel-Plana formula,
we rewrite this infinite series as a sum of two integrals.
We examine the physical meanings of these two integrals
and find that the first one represents the initial collapse (the semi-classical limit)
and the second one represents the revival (the quantum correction) in the JCM.
Furthermore,
we evaluate the first and second-order perturbations
for the time-evolution of the JCM with an initial thermal coherent state
for the cavity mode at low temperature,
and write down their correction terms
as sums of integrals by making use of the Abel-Plana formula.
\end{abstract}

\section{\label{section-Introduction}Introduction}

The Jaynes-Cummings model (JCM),
which described the interaction between a two-level atom and a single electromagnetic field mode,
was originally proposed for examining spontaneous emission in 1960s~\cite{Jaynes1963}.
This model is derived from applying
the rotating wave approximation
to an electric dipole coupling.
In the interaction term of the Hamiltonian of the JCM,
the photon creation operator accompanies the atomic de-excitation operator
and the photon annihilation operator accompanies the atomic excitation operator.
Because the JCM is an exactly soluble quantum mechanical model,
it is investigated theoretically by researchers
in the field of quantum optics eagerly~\cite{Louisell1973,Shore1993,Schleich2001}.

If we initially prepare the atom in the ground state and the cavity mode in a coherent state,
the JCM causes the collapse and the revival of the Rabi oscillations many times
in a complicated pattern in its time-evolution
and this phenomenon is regarded
as the evidence of the quantum nature of the electromagnetic field~\cite{Cummings1965,Eberly1980}.
(This phenomenon was confirmed experimentally in 1980s~\cite{Rempe1987}.)
Thus, the demonstration of the collapse and the revival of the Rabi oscillations in the JCM gives
the foundation to Planck's thought~\cite{Schiff1968}.
That is, the collapse and the revival of the Rabi oscillations in the JCM
tells us that the photon's energy is equal to $h\nu$,
where $h$ represents the Planck's constant and $\nu$ represents the frequency of the photon,
so that excitation of the photons shows discreteness.

Recently, the JCM has been studied from a new viewpoint
by the researchers in the field of quantum information science.
The JCM is often used for investigating the evolution of entanglement
between the atom and the single mode cavity field~\cite{Bose2001,Scheel2003}.
The lower bound of entanglement between the two-level atom and the thermal photons
in the JCM is also discussed~\cite{Azuma2008-PRA}.
The JCM can be applied to the realization of quantum computation~\cite{Azuma2008-Preprint}.
So-called sudden death effect
(disappearance of entanglement of two isolated Jaynes-Cummings atoms
in a finite time)
is predicted~\cite{Yu2004,Yonac2006,Yu2006},
and it is experimentally demonstrated~\cite{Almeida2007}.
Thus, some researchers in the field of quantum information science think
that the JCM has to be studied from the new viewpoint.

When we discuss the JCM,
we often have to handle an intractable infinite series.
(If the electromagnetic field is resonant with the atom,
the $n$th term of this infinite series is given by a trigonometric function for $\sqrt{n}t$,
where $t$ represents a variable of the time.)
For example, the atomic population inversion in the collapse and the revival
of the Rabi oscillations in the JCM is described by this infinite series.
In the thermal JCM, whose initial state of the cavity mode is given by a thermal equilibrium state,
the atomic population inversion is written as a similar intractable infinite series, as well.

In Ref.~\cite{Klimov1999},
Klimov and Chumakov discuss the thermal JCM and evaluate the atomic population inversion,
which is described by the intractable infinite series.
They change this intractable infinite series into a sum of two integrals
by making use of the Abel-Plana formula mentioned in Ref.~\cite{Whittaker1927}.
By carrying out numerical calculations,
they find that the first integral represents a semi-classical limit (the initial collapse)
and the second integral represents a quantum correction (quasi-chaotic behavior).

In this paper, we give an analytical treatment to study the behavior
of the collapse and the revival of the Rabi oscillations in the JCM,
according to the method proposed by Klimov and Chumakov.
For applying the Abel-Plana formula to the infinite series
that describes the atomic population inversion,
we replace an inverse of a factorial $1/n!$
with an inverse of the gamma function $1/\Gamma(n+1)$
and perform the analytical continuation on the complex plane as $1/\Gamma(z+1)$.
(This prescription is a new key point of this paper as compared with
Ref.~\cite{Klimov1999}.)
After giving this step,
using the Abel-Plana formula,
we write down the atomic population inversion as a sum of two integrals.
We examine the physical meaning of these two integrals and find that
the first integral represents the initial collapse (the semi-classical limit)
and the second integral represents the revival (the quantum correction) in the JCM.
In this paper, we clarify that we can separate the quantum correction
from the semi-classical limit in the solution of the JCM.

Furthermore, we evaluate the first and second-order perturbations
for the
\\
time-evolution of the JCM,
whose initial state of the cavity mode is given
by a thermal coherent state at low temperature~\cite{Ojima1981,Mann1989}.
[The expansion parameter for the perturbation theory is given by
$\theta(\beta)\simeq\exp(-\beta\epsilon/2)$,
where $\beta=(k_{\mbox{\scriptsize B}}T)^{-1}$,
$\epsilon=h\nu$,
$\beta\epsilon\gg 1$
and $\nu$ is the frequency of the cavity field.
A rigorous definition of $\theta(\beta)$ is given by Refs.~\cite{Ojima1981} and \cite{Mann1989}.]
We obtain the first and second-order correction terms of the atomic population inversion
and rewrite them as sums of integrals,
using Klimov and Chumakov's method and the Abel-Plana formula.

After deriving integral forms of the atomic population inversion
for both resonant and off-resonant cases
and their thermal perturbation corrections,
we examine effects of detuning and low temperature
against the collapse and the revival of the Rabi oscillations numerically.

This paper is organized as follows.
In section~\ref{section-JCM-collapse-revival},
we give a review of the JCM and its collapse and revival of the Rabi oscillations.
In section~\ref{section-Representing-atomic-population-inversion},
we apply Klimov and Chumakov's method to the atomic population inversion,
which is represented by the intractable infinite series.
(In this section, we consider the zero-temperature case.)
We rewrite it as a sum of two integrals with the Abel-Plana formula.
In section~\ref{section-thermal-coherent-state},
we consider the time-evolution of the JCM at low temperature.
Preparing the cavity field in a thermal coherent state initially,
we evaluate the first and second-order perturbation corrections
of the atomic population inversion.
Using the Abel-Plana formula,
we rewrite them as sums of integrals.
In section~\ref{section-Properties-integrals},
we examine properties of the integrals obtained
in sections~\ref{section-Representing-atomic-population-inversion}
and \ref{section-thermal-coherent-state}
in detail by numerical calculations and analytical methods.
In section~\ref{section-Discussions},
we give a brief discussion.
In Appendix~\ref{appendix-Abel-Plana-formula},
we give a derivation of the Abel-Plana formula,
which plays an important role in Klimov and Chumakov's method.
In Appendix~\ref{appendix-remarks-numerical-calculations},
we give some remarks about precise techniques
for carrying out the numerical calculations shown in section~\ref{section-Properties-integrals}.

\section{\label{section-JCM-collapse-revival}
The JCM and its collapse and revival of the Rabi oscillations}

The JCM is a quantum system, which consists of the single two-level atom
and the single mode of the electromagnetic field.
Its Hamiltonian is given by
\begin{equation}
H
=
\frac{\hbar}{2}\omega_{0}\sigma_{z}
+
\hbar\omega a^{\dagger}a
+
\hbar\kappa(\sigma_{+}a+\sigma_{-}a^{\dagger}),
\label{JCM-Hamiltonian-0}
\end{equation}
where
$\hbar=h/2\pi$,
$\sigma_{\pm}=(1/2)(\sigma_{x}\pm i\sigma_{y})$
and
$[a,a^{\dagger}]=1$.
The Pauli matrices
($\sigma_{i}$, $i=x,y,z$)
are operators of the atom,
and $a$ and $a^{\dagger}$ are an annihilation and a creation operators of photons, respectively.
Moreover, we assume the coupling constant $\kappa$ to be real.

We can divide $H$
given by Eq.~(\ref{JCM-Hamiltonian-0}) as follows:
\begin{eqnarray}
H&=&\hbar(C_{1}+C_{2}), \nonumber \\
C_{1}&=&\omega[(1/2)\sigma_{z}+a^{\dagger}a], \nonumber \\
C_{2}&=&\kappa(\sigma_{+}a+\sigma_{-}a^{\dagger})-(\Delta\omega/2)\sigma_{z},
\end{eqnarray}
where
$\Delta\omega=\omega-\omega_{0}$.
Because $[C_{1},C_{2}]=0$ and
we can diagonalize $C_{1}$ at ease,
we take the following interaction picture.
We write a state vector of the whole system in the Schr{\"o}dinger picture as
$|\psi_{\mbox{\scriptsize S}}(t)\rangle$.
We define a state vector in the interaction picture as
$|\psi_{\mbox{\scriptsize I}}(t)\rangle
=\exp(iC_{1}t)|\psi_{\mbox{\scriptsize S}}(t)\rangle$.
[We assume $|\psi_{\mbox{\scriptsize I}}(0)\rangle=|\psi_{\mbox{\scriptsize S}}(0)\rangle$.]
The time-evolution of $|\psi_{\mbox{\scriptsize I}}(t)\rangle$ is given by
$|\psi_{\mbox{\scriptsize I}}(t)\rangle=U(t)|\psi_{\mbox{\scriptsize I}}(0)\rangle$,
where $U(t)=\exp(-iC_{2}t)$.

We give the basis vectors for the state of the atom and the photons of the cavity field as follows.
(In this section, we consider only the zero-temperature case.)
We write the ground and excited states of the atom as two-component vectors,
\begin{equation}
|g\rangle_{\mbox{\scriptsize A}}
=
\left(
\begin{array}{c}
0 \\
1
\end{array}
\right),
\quad\quad
|e\rangle_{\mbox{\scriptsize A}}
=
\left(
\begin{array}{c}
1 \\
0
\end{array}
\right),
\label{atom-basis-vectors}
\end{equation}
where we assume that $|g\rangle_{\mbox{\scriptsize A}}$ and $|e\rangle_{\mbox{\scriptsize A}}$ are
eigenvectors of $\sigma_{z}$
with eigenvalues $-1$ and $1$, respectively.
(The index A stands for the atom.)
We describe the number states of the photons as
$|n\rangle_{\mbox{\scriptsize P}}$ ($n=0,1,2,...$),
which are eigenstates of $a^{\dagger}a$.
(The index P stands for the photons.)

Describing the atom's Pauli operators by $2\times 2$ matrices,
we can write down $U(t)$ as follows:
\begin{equation}
U(t)
=
\exp[-it
\left(
\begin{array}{cc}
-\Delta\omega/2 & \kappa a \\
\kappa a^{\dagger} & \Delta\omega/2
\end{array}
\right)
]
=
\left(
\begin{array}{cc}
u_{00} & u_{01} \\
u_{10} & u_{11}
\end{array}
\right),
\label{unitary-evolution-1}
\end{equation}
where
\begin{eqnarray}
u_{00}
&=&
\cos(\sqrt{D+\kappa^{2}}t)
-\frac{i}{2}\Delta\omega
\frac{\sin(\sqrt{D+\kappa^{2}}t)}{\sqrt{D+\kappa^{2}}}, \nonumber \\
u_{01}
&=&
i\kappa
\frac{\sin(\sqrt{D}t)}{\sqrt{D}}
a, \nonumber \\
u_{10}
&=&
i\kappa
\frac{\sin(\sqrt{D+\kappa^{2}}t)}{\sqrt{D+\kappa^{2}}}
a^{\dagger}, \nonumber \\
u_{11}
&=&
\cos(\sqrt{D}t)
+\frac{i}{2}\Delta\omega
\frac{\sin(\sqrt{D}t)}{\sqrt{D}},
\label{unitary-evolution-2} 
\end{eqnarray}
and
\begin{equation}
D
=
(\Delta\omega/2)^{2}+\kappa^{2}a^{\dagger}a.
\label{unitary-evolution-3}
\end{equation}

Here, we define the initial state of the whole system as follows.
We assume that the atom is in the ground state $|g\rangle_{\mbox{\scriptsize A}}$ at $t=0$.
Moreover, we assume that the cavity mode is in the coherent state
$|\alpha\rangle_{\mbox{\scriptsize P}}$ at $t=0$, where
\begin{equation}
|\alpha\rangle_{\mbox{\scriptsize P}}
=
\exp(-|\alpha|^{2}/2)\sum_{n=0}^{\infty}\frac{\alpha^{n}}{\sqrt{n!}}
|n\rangle_{\mbox{\scriptsize P}},
\label{definition-coherent-light-1}
\end{equation}
and $\alpha$ is an arbitrary complex number.
(We provide that $0!=1$.)
From now on, to let the discussion be simple,
we assume $\alpha$ to be real.
Writing the initial state of the whole system as
$|\psi_{\mbox{\scriptsize I}}(0)\rangle
=|g\rangle_{\mbox{\scriptsize A}}|\alpha\rangle_{\mbox{\scriptsize P}}$,
the time-evolution $|\psi_{\mbox{\scriptsize I}}(t)\rangle$ is given as follows:
\begin{equation}
|\psi_{\mbox{\scriptsize I}}(t)\rangle
=
U(t)|\psi_{\mbox{\scriptsize I}}(0)\rangle
=
\left(
\begin{array}{cc}
u_{00} & u_{01} \\
u_{10} & u_{11}
\end{array}
\right)
\left(
\begin{array}{c}
0 \\
|\alpha\rangle_{\mbox{\scriptsize P}}
\end{array}
\right)
=
\left(
\begin{array}{c}
u_{01}|\alpha\rangle_{\mbox{\scriptsize P}} \\
u_{11}|\alpha\rangle_{\mbox{\scriptsize P}}
\end{array}
\right).
\end{equation}
Thus, the probability that we observe $|g\rangle_{\mbox{\scriptsize A}}$
at the time $t$ is given by
\begin{eqnarray}
P_{g}(t)
&=&
|_{\mbox{\scriptsize A}}\langle g|\psi_{\mbox{\scriptsize I}}(t)\rangle|^{2} \nonumber \\
&=&
|u_{11}|\alpha\rangle_{\mbox{\scriptsize P}}|^{2} \nonumber \\
&=&
\exp(-\alpha^{2})
\sum_{n=0}^{\infty}
\frac{\alpha^{2n}}{n!}
[\cos^{2}(\sqrt{(\Delta\omega/2)^{2}+n\kappa^{2}}t) \nonumber \\
&&
\quad
+(\Delta\omega/2)^{2}
\frac{\sin^{2}(\sqrt{(\Delta\omega/2)^{2}+n\kappa^{2}}t)}{(\Delta\omega/2)^{2}+n\kappa^{2}}].
\label{probability-ground-state-01}
\end{eqnarray}

\begin{figure}
\begin{center}
\includegraphics[scale=1.0]{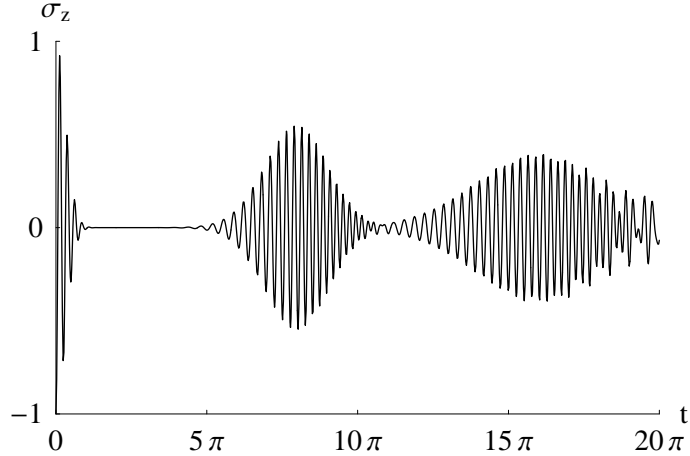}
\end{center}
\caption{A graph of $\langle \sigma_{z}(t)\rangle$ ($t\in[0,20\pi]$)
for $\kappa=1$, $\Delta\omega=0$ and $\alpha=4$.
In the numerical calculations of $\langle \sigma_{z}(t)\rangle$
defined in Eq.~(\ref{atomic-population-inversion-equation01}),
the summation of the index $n$ is carried out up to $n=100$.
Looking at this graph, we notice that the initial collapse time is order of unity
and the period of the revival is approximately equal to $8\pi$.}
\label{Figure01}
\end{figure}

The atomic population inversion is given by
\begin{equation}
\langle \sigma_{z}(t)\rangle
=
\mbox{Tr}_{\mbox{\scriptsize P}}
\langle \psi_{\mbox{\scriptsize I}}(t)|
\sigma_{z}|\psi_{\mbox{\scriptsize I}}(t)\rangle
=
1-2P_{g}(t).
\label{atomic-population-inversion-equation00}
\end{equation}
Especially, in the case where $\Delta\omega=0$,
that is,
the electromagnetic field is resonant with the atom
so that the energy gap is equal to the frequency of the photons,
we can replace $|\kappa|t$ with $t$ by letting the time $t$ be in units of $|\kappa|^{-1}$
and we obtain
\begin{equation}
\langle \sigma_{z}(t)\rangle
=
-\exp(-\alpha^{2})
\sum_{n=0}^{\infty}\frac{\alpha^{2n}}{n!}\cos(2\sqrt{n}t).
\label{atomic-population-inversion-equation01}
\end{equation}
We plot $\langle \sigma_{z}(t)\rangle$ ($t\in[0,20\pi]$)
for $\kappa=1$, $\Delta\omega=0$ and $\alpha=4$
in Fig.~\ref{Figure01}.

Looking at Fig.~\ref{Figure01},
we can observe the collapse and the revival of the Rabi oscillations clearly.
In general, the larger $|\alpha|$ is,
the more distinctly we observe the revival of the Rabi oscillations.
From Fig.~\ref{Figure01},
we can suppose that the time scale of the initial collapse is order of unity and
the period of the revival is around $2\pi|\alpha|$.

The time scale of the initial collapse and the period of the revival
are explained as follows~\cite{Walls1994,Barnett1997}.
Let us evaluate $\langle \sigma_{z}(t)\rangle$ defined
in Eq.~(\ref{atomic-population-inversion-equation01})
with assuming $\alpha^{2}\gg 1$.
Writing the index of the summation as $n=\alpha^{2}+\delta n$,
because of the property of the Poisson distribution,
the major contribution to the right-hand side of Eq.~(\ref{atomic-population-inversion-equation01})
comes from the terms with $|\delta n|<\alpha^{2}$,
so that we can neglect the terms with $|\delta n|\geq \alpha^{2}$.
We rewrite $\sqrt{n}$ as
\begin{equation}
\sqrt{n}
\simeq
\frac{\alpha^{2}+n}{2|\alpha|}.
\end{equation}
Then, we can obtain an approximate form $\langle \sigma_{z}(t)\rangle$
in Eq.~(\ref{atomic-population-inversion-equation01})
as
\begin{eqnarray}
\langle \sigma_{z}(t)\rangle
&\simeq &
-\exp(-\alpha^{2})
\sum_{n=0}^{\infty}\frac{\alpha^{2n}}{n!}
\frac{1}{2}
[\exp(2i\frac{\alpha^{2}+n}{2|\alpha|}t)
+
\exp(-2i\frac{\alpha^{2}+n}{2|\alpha|}t)] \nonumber \\
&=&
-\exp[\alpha^{2}(\cos\frac{t}{|\alpha|}-1)]
\cos(|\alpha|t+\alpha^{2}\sin\frac{t}{|\alpha|}).
\label{atomic-population-inversion-approximation02}
\end{eqnarray}

In the right-hand side of Eq.~(\ref{atomic-population-inversion-approximation02}),
$\exp[\alpha^{2}(\cos(t/|\alpha|)-1)]$ represents an amplitude envelope of the wave,
which causes a kind of the beat,
and $\cos[|\alpha|t+\alpha^{2}\sin(t/|\alpha|)]$ represents the Rabi oscillations.
If we assume $0\leq t/|\alpha|\ll 1$,
we obtain
$\cos(t/|\alpha|)\simeq 1-(t^{2}/2\alpha^{2})$,
and
the factor of the amplitude envelope approximates to
$\exp[\alpha^{2}(\cos(t/|\alpha|)-1)]\sim \exp(-t^{2}/2)$,
so that we understand that the initial collapse time is approximately equal to unity.
Moreover, the factor of the amplitude envelope
$\exp[\alpha^{2}(\cos(t/|\alpha|)-1)]$ shows that the period of the revival
is given by $2\pi|\alpha|$.

Thus, Eq.~(\ref{atomic-population-inversion-approximation02})
represents the collapse time and the period of the revival of the genuine
$\langle \sigma_{z}(t)\rangle$
given by Eq.~(\ref{atomic-population-inversion-equation01}) well.
However, to investigate properties of $\langle \sigma_{z}(t)\rangle$ precisely,
changing the infinite series that appears
in Eqs.~(\ref{probability-ground-state-01}) and (\ref{atomic-population-inversion-equation00})
into a simple form is favorable.
This is the motivation of this paper.

\section{\label{section-Representing-atomic-population-inversion}
Representing the atomic population inversion as a sum of two integrals
for the zero-temperature case}

To rewrite the atomic population inversion $\langle \sigma_{z}(t)\rangle$
given by Eqs.~(\ref{probability-ground-state-01})
and (\ref{atomic-population-inversion-equation00}),
we use the following formula.
If $n_{1}$ and $n_{2}$ are integers and $\phi(z)$ is a function
which is analytical and bounded for all complex values of $z$ such that
$n_{1}\leq\mbox{Re}(z)\leq n_{2}$, then
\begin{eqnarray}
&&
\frac{1}{2}\phi(n_{1})
+\phi(n_{1}+1)
+\phi(n_{1}+2)
+...
+\phi(n_{2}-1)
+\frac{1}{2}\phi(n_{2})  \nonumber \\
&=&
\int_{n_{1}}^{n_{2}}\phi(x)dx
-i
\int_{0}^{\infty}
\frac{1}{e^{2\pi y}-1} \nonumber\\
&&
\times
[
\phi(n_{2}+iy)-\phi(n_{1}+iy)-\phi(n_{2}-iy)+\phi(n_{1}-iy)
]dy.
\label{Abel-Plana-formula-0}
\end{eqnarray}
(This formula is described in Ref.~\cite{Whittaker1927} as an example.
The author of Ref.~\cite{Whittaker1927} mentions only a suggestion to prove this formula.
We give details of the derivation of this formula in Appendix~\ref{appendix-Abel-Plana-formula}.)
Moreover, if we assume $\phi(z)\to 0$ as $\mbox{Re}(z)\to +\infty$,
we obtain
\begin{equation}
\frac{1}{2}\phi(0)
+
\sum_{n=1}^{\infty}
\phi(n)
=
\int_{0}^{\infty}
\phi(x)dx
+i
\int_{0}^{\infty}
\frac{\phi(iy)-\phi(-iy)}{e^{2\pi y}-1}
dy.
\label{Abel-Plana-formula-1}
\end{equation}
This equation is called the Abel-Plana formula.

The Abel-Plana formula given by Eq.~(\ref{Abel-Plana-formula-1})
changes an infinite series into a sum of two integrals.
However, we cannot apply this formula to $\langle \sigma_{z}(t)\rangle$
given by Eqs.~(\ref{probability-ground-state-01})
and (\ref{atomic-population-inversion-equation00})
in a straightforward manner.
Thus, we try to extend this formula in the following way.

First, we consider a series,
\begin{equation}
\sum_{n=0}^{\infty}
\frac{1}{n!}f(n+c),
\end{equation}
where $c$ is a real constant and the function $f(x)$ is infinitely differentiable
and bounded for any real value of $x$.
Thus, we can rewrite $f(n+c)$ as the Taylor series and we obtain
\begin{equation}
\sum_{n=0}^{\infty}
\frac{1}{n!}f(n+c)
=
\sum_{n=0}^{\infty}\frac{1}{n!}
\sum_{m=0}^{\infty}\frac{c^{m}}{m!}
f^{(m)}(n).
\label{Taylor-series-01}
\end{equation}
[In Eq.~(\ref{Taylor-series-01}), $f^{(m)}(n)$ denotes
$(d^{m}/dx^{m})f(x)|_{x=n}$.]

Second, using the property of the gamma function
$\Gamma(n+1)=n!$,
we rewrite Eq.~(\ref{Taylor-series-01}) as
\begin{equation}
\sum_{n=0}^{\infty}
\frac{1}{n!}f(n+c)
=
\sum_{m=0}^{\infty}\frac{c^{m}}{m!}
\sum_{n=0}^{\infty}\frac{1}{\Gamma(n+1)}
f^{(m)}(n), \label{Taylor-series-02}
\end{equation}
where
\begin{equation}
\Gamma(z)=\int_{0}^{\infty}t^{z-1}e^{-t}dt.
\label{definition-gamma-function-0}
\end{equation}
The gamma function $\Gamma(z)$ given by Eq.~(\ref{definition-gamma-function-0})
converges absolutely only for $\mbox{Re}(z)>0$.
However, by analytical continuation,
we can let $\Gamma(z)$ be analytical everywhere
in the complex plane except at $z=0,-1,-2,...$.
Because there are no points at which $\Gamma(z)$ is equal to zero,
$1/\Gamma(z)$ is analytical at all finite points of the complex plane.

Third, we assume that
$f^{(m)}(z)/\Gamma(z+1)$ is analytical and bounded for all complex values of $z$
such that $0\leq \mbox{Re}(z)<+\infty$
and $f^{(m)}(z)/\Gamma(z+1)\rightarrow 0$ as $\mbox{Re}(z)\rightarrow +\infty$ for $m=0,1,2,...$.
Then, we can apply the Abel-Plana formula given by Eq.~(\ref{Abel-Plana-formula-1})
to the right-hand side of Eq.~(\ref{Taylor-series-02})
and we obtain
\begin{eqnarray}
\sum_{n=0}^{\infty}
\frac{1}{n!}f(n+c)
&=&
\sum_{m=0}^{\infty}\frac{c^{m}}{m!}
\Bigl[
\frac{1}{2}f^{(m)}(0)
+
\int_{0}^{\infty}
\frac{f^{(m)}(x)}{\Gamma(x+1)}dx \nonumber \\
&&\quad
+
i\int_{0}^{\infty}\frac{1}{e^{2\pi y}-1}
[\frac{f^{(m)}(iy)}{\Gamma(1+iy)}-\frac{f^{(m)}(-iy)}{\Gamma(1-iy)}]
dy
\Bigr] \nonumber \\
&=&
\frac{1}{2}f(c)
+
\int_{0}^{\infty}
\frac{f(x+c)}{\Gamma(x+1)}
dx \nonumber \\
&&\quad
-2
\int_{0}^{\infty}
\frac{1}{e^{2\pi y}-1}
\mbox{Im}
\{
\frac{f(c+iy)}{\Gamma(1+iy)}
\}dy. \label{modified-Abel-Plana-formula-01}
\end{eqnarray}

Here, applying the formula given by Eq.~(\ref{modified-Abel-Plana-formula-01})
to Eq.~(\ref{probability-ground-state-01}),
we can change $P_{g}(t)$ represented as the infinite series
into a sum of integrals.
At first,
we define
\begin{equation}
c=(\Delta\omega/2\kappa)^{2}.
\label{definition-constant-c}
\end{equation}
Then, defining $f(z+c)$ as
\begin{equation}
f(z+c)
=
|\alpha|^{2(z+c)}
[\cos^{2}(\sqrt{z+c}|\kappa|t)
+\frac{c}{z+c}\sin^{2}(\sqrt{z+c}|\kappa|t)],
\end{equation}
we can confirm that $f(z+c)/\Gamma(z+1)$ is analytical and bounded
for all complex plane.
Moreover, $f(z+c)/\Gamma(z+1)\rightarrow 0$ as $\mbox{Re}(z)\rightarrow +\infty$.
[In the limit of $\mbox{Re}(z)\rightarrow +\infty$,
$|\Gamma(z+1)|$ increases more rapidly than any exponential functions of $z$.]
Thus,
we obtain
\begin{eqnarray}
P_{g}(t)
&=&
\exp(-\alpha^{2})|\alpha|^{-2c}
\sum_{n=0}^{\infty}
\frac{|\alpha|^{2(n+c)}}{n!}
[
\cos^{2}(\sqrt{n+c}|\kappa|t)
+
\frac{c}{n+c}\sin^{2}(\sqrt{n+c}|\kappa|t)] \nonumber \\
&=&
\exp(-\alpha^{2})
[(1/2)+I^{(0)}_{1}(t)-2I^{(0)}_{2}(t)],
\label{Pgt-integral-form}
\end{eqnarray}
where
\begin{eqnarray}
I^{(l)}_{1}(t)
&=&
\int_{0}^{\infty}
\frac{|\alpha|^{2x}}{\Gamma(x+1)}
[\cos^{2}(\sqrt{x+c+l}|\kappa|t)
+\frac{c}{x+c+l}\sin^{2}(\sqrt{x+c+l}|\kappa|t)]dx. \nonumber \\
I^{(l)}_{2}(t)
&=&
\int_{0}^{\infty}
\frac{1}{e^{2\pi y}-1}
\mbox{Im}\{
\frac{|\alpha|^{2iy}}{\Gamma(1+iy)}
[\cos^{2}(\sqrt{c+l+iy}|\kappa|t) \nonumber \\
&&\quad
+\frac{c}{c+l+iy}\sin^{2}(\sqrt{c+l+iy}|\kappa|t)]\}dy.
\label{definition-Il1t-Il2t-i}
\end{eqnarray}
In the above equation, $P_{g}(t)$ is represented
as a sum of two integrals,
$\int_{0}^{\infty}dx$ and $\int_{0}^{\infty}dy$
[namely, $I^{(0)}_{1}(t)$ and $I^{(0)}_{2}(t)$].
Thus, using Eq.~(\ref{atomic-population-inversion-equation00}),
we can describe $\langle\sigma_{z}(t)\rangle$ as a sum of these integrals,
\begin{equation}
\langle\sigma_{z}(t)\rangle
=
1-\exp(-\alpha^{2})
[1+2I^{(0)}_{1}(t)-4I^{(0)}_{2}(t)].
\end{equation}

In the case where $\Delta\omega=0$, $\kappa=1$ and $c=0$,
we obtain
\begin{eqnarray}
P_{g}(t)
&=&
\exp(-\alpha^{2})
[
\frac{1}{2}
+
\int_{0}^{\infty}
\frac{|\alpha|^{2x}}{\Gamma(x+1)}
\cos^{2}(\sqrt{x}t)
dx \nonumber \\
&&\quad
-2
\int_{0}^{\infty}
\frac{1}{e^{2\pi y}-1}
\mbox{Im}\{
\frac{|\alpha|^{2iy}}{\Gamma(1+iy)}
\cos^{2}(\sqrt{iy}t)
\}dy
] \nonumber \\
&=&
\frac{1}{2}
+
\exp(-\alpha^{2})
[
\frac{1}{4}
+
\frac{1}{2}
\int_{0}^{\infty}
\frac{|\alpha|^{2x}}{\Gamma(x+1)}
\cos(2\sqrt{x}t)
dx \nonumber \\
&&\quad
-
\int_{0}^{\infty}
\frac{1}{e^{2\pi y}-1}
\mbox{Im}\{
\frac{|\alpha|^{2iy}}{\Gamma(1+iy)}
\cos(2\sqrt{iy}t)
\}dy
],
\end{eqnarray}
where we use
\begin{equation}
\frac{1}{2}
\int_{0}^{\infty}
\frac{|\alpha|^{2x}}{\Gamma(x+1)}
dx
-
\int_{0}^{\infty}
\frac{1}{e^{2\pi y}-1}
\mbox{Im}\{
\frac{|\alpha|^{2iy}}{\Gamma(1+iy)}
\}dy
=
-\frac{1}{4}+\frac{1}{2}\exp(\alpha^{2}).
\label{integral-formula-derived-from-Abel-Plana-A}
\end{equation}
[We can derive Eq.~(\ref{integral-formula-derived-from-Abel-Plana-A})
from Eq.~(\ref{Abel-Plana-formula-1}) at ease.]
Thus, we obtain
\begin{equation}
\langle \sigma_{z}(t)\rangle
=
-(1/2)\exp(-\alpha^{2})
+J_{1}(t)
+J_{2}(t),
\label{atomic-population-inversion-equation03}
\end{equation}
where
\begin{eqnarray}
J_{1}(t)
&=&
-\exp(-\alpha^{2})
\int_{0}^{\infty}
\frac{|\alpha|^{2x}}{\Gamma(x+1)}\cos(2\sqrt{x}t)dx, \nonumber \\
J_{2}(t)
&=&
2\exp(-\alpha^{2})
\int_{0}^{\infty}
\frac{1}{e^{2\pi y}-1}
\mbox{Im}
\{
\frac{|\alpha|^{2iy}}{\Gamma(1+iy)}\cos(2\sqrt{iy}t)
\}dy.
\label{definition-first-and-second-integral00}
\end{eqnarray}

\section{\label{section-thermal-coherent-state}
The time-evolution of the JCM with an initial thermal coherent state}

\subsection{\label{subsection-definition-thermal-coherent-state}
The definition of the thermal coherent state}

In Refs.~\cite{Ojima1981} and \cite{Mann1989},
a thermal coherent state is defined as an extension of the zero-temperature coherent state
according to the thermo field dynamics.
In this section,
preparing the atom in the ground state
and the cavity mode in the thermal coherent state initially,
and letting the whole system evolve in time with the JCM,
we calculate the atomic population inversion
up to the second-order perturbation correction.
The obtained first and second-order correction terms
are represented as the intractable infinite series.
In subsection~\ref{subsection-atomic-population-inversion-second-order-perturbation},
we rewrite these correction terms as sums of integrals
by making use of the Abel-Plana formula.

In the thermo field dynamics,
we have to handle a space that is a direct product
of the ordinary zero-temperature Hilbert space ${\cal H}$
and a so-called tilde space $\tilde{{\cal H}}$.
Thus, every number state of the photons $|n\rangle\in{\cal H}$
has a corresponding state $|\tilde{n}\rangle\in\tilde{{\cal H}}$,
so that the orthogonal basis for the whole space
in thermo field dynamics is described as
$\{|n\rangle\otimes|\tilde{m}\rangle
\in{\cal H}\otimes\tilde{{\cal H}}:
n,m\in\{0,1,2,...\}\}$.
In the next paragraph, we give a definition of the thermal coherent state.

First of all, we define creation and annihilation operators
acting on ${\cal H}$ and $\tilde{{\cal H}}$ as follows:
\begin{equation}
[a,a^{\dagger}]=[\tilde{a},\tilde{a}^{\dagger}]=1,
\quad
[a,\tilde{a}]=[a,\tilde{a}^{\dagger}]=0.
\end{equation}
We pay attention to the fact that the operators on ${\cal H}$
(namely $a$ and $a^{\dagger}$)
and the operators on $\tilde{{\cal H}}$
(namely $\tilde{a}$ and $\tilde{a}^{\dagger}$)
commute with each other.
Next, we introduce the temperature by the following unitary transformation:
\begin{equation}
\hat{U}_{\beta}=\exp(-i\hat{G}),
\end{equation}
\begin{equation}
\hat{G}=-i\theta(\beta)(a^{\dagger}\tilde{a}^{\dagger}-a\tilde{a}),
\end{equation}
\begin{eqnarray}
\cosh\theta(\beta)&=&[1-\exp(-\beta\epsilon)]^{-1/2}, \nonumber \\
\sinh\theta(\beta)&=&[\exp(\beta\epsilon)-1]^{-1/2},
\label{definition-theta-beta-1}
\end{eqnarray}
where $\beta=(k_{\mbox{\scriptsize B}}T)^{-1}$ and $\epsilon=\hbar\omega$.
We note that $\theta(\beta)$ is real.
To emphasize that $\hat{G}$ and $\hat{U}_{\beta}$ are operators
acting on both ${\cal H}$ and $\tilde{{\cal H}}$,
we put an accent (a hat) on them.
Then, annihilation operators are transformed as follows:
\begin{eqnarray}
a&\rightarrow&a(\beta)=\hat{U}_{\beta}a\hat{U}_{\beta}^{\dagger}, \nonumber \\
\tilde{a}&\rightarrow&\tilde{a}(\beta)=\hat{U}_{\beta}\tilde{a}\hat{U}_{\beta}^{\dagger}.
\label{temperature-transformations-a-tilde-a}
\end{eqnarray}
Writing down the zero-temperature vacuum as
$|0\rangle\otimes|\tilde{0}\rangle=|0,\tilde{0}\rangle\in{\cal H}\otimes\tilde{{\cal H}}$,
the thermal vacuum is given by
\begin{equation}
|0(\beta)\rangle=\hat{U}_{\beta}|0,\tilde{0}\rangle.
\label{definition-thermal-vacuum}
\end{equation}
Then, a thermal coherent state is defined as follows:
\begin{equation}
|\alpha,\tilde{\gamma};\beta\rangle
=
\exp[\alpha a^{\dagger}(\beta)
+\tilde{\gamma}^{*}\tilde{a}^{\dagger}(\beta)
-\alpha^{*}a(\beta)
-\tilde{\gamma}\tilde{a}(\beta)]
|0(\beta)\rangle,
\label{definition-thermal-coherent-state-1}
\end{equation}
where $\alpha$ and $\tilde{\gamma}$ are arbitrary complex numbers.
From now on,
for the simplicity,
we assume $\alpha$ and $\tilde{\gamma}$ to be real.

Writing the initial state of the whole system as
$|\psi_{\mbox{\scriptsize I}}(0)\rangle
=|g\rangle_{\mbox{\scriptsize A}}
|\alpha,\tilde{\gamma};\beta\rangle_{\mbox{\scriptsize P}}$,
the time-evolution of $|\psi_{\mbox{\scriptsize I}}(t)\rangle$ is described as
\begin{equation}
|\psi_{\mbox{\scriptsize I}}(t)\rangle
=
\left(
\begin{array}{l}
u_{01}|\alpha,\tilde{\gamma};\beta\rangle_{\mbox{\scriptsize P}} \\
u_{11}|\alpha,\tilde{\gamma};\beta\rangle_{\mbox{\scriptsize P}}
\end{array}
\right),
\end{equation}
where $u_{01}$ and $u_{11}$ are defined in Eqs.~(\ref{unitary-evolution-2})
and (\ref{unitary-evolution-3}).
Thus, the probability that we observe $|g\rangle_{\mbox{\scriptsize A}}$
at the time $t$ is given by
\begin{equation}
P_{g}(\beta;t)=|u_{11}|\alpha,\tilde{\gamma};\beta\rangle|^{2}.
\label{probability-ground-state-thermal-1}
\end{equation}
To calculate $P_{g}(\beta;t)$ by the perturbation theory,
we rewrite $|\alpha,\tilde{\gamma};\beta\rangle$
given by Eqs.~(\ref{temperature-transformations-a-tilde-a}),
(\ref{definition-thermal-vacuum})
and (\ref{definition-thermal-coherent-state-1})
as follows:
\begin{eqnarray}
|\alpha,\tilde{\gamma};\beta\rangle
&=&
\exp[
\hat{U}_{\beta}
(\alpha a^{\dagger}
+\tilde{\gamma}\tilde{a}^{\dagger}
-\alpha a
-\tilde{\gamma}\tilde{a})
\hat{U}_{\beta}^{\dagger}
]
\hat{U}_{\beta}
|0,\tilde{0}\rangle \nonumber \\
&=&
\hat{U}_{\beta}
\exp[
\alpha a^{\dagger}
+\tilde{\gamma}\tilde{a}^{\dagger}
-\alpha a
-\tilde{\gamma}\tilde{a}
]
|0,\tilde{0}\rangle \nonumber \\
&=&
\hat{U}_{\beta}
\exp[\alpha(a^{\dagger}-a)]
\exp[\tilde{\gamma}(\tilde{a}^{\dagger}-\tilde{a}]
|0,\tilde{0}\rangle \nonumber \\
&=&
\hat{U}_{\beta}|\alpha\rangle|\tilde{\gamma}\rangle,
\end{eqnarray}
where $|\alpha\rangle$ and $|\tilde{\gamma}\rangle$ represent the (zero-temperature)
coherent states in ${\cal H}$ and $\tilde{{\cal H}}$, respectively.
Thus, we can change Eq.~(\ref{probability-ground-state-thermal-1})
into the following form:
\begin{equation}
P_{g}(\beta;t)
=
\langle \alpha|\langle\tilde{\gamma}|
\hat{U}_{\beta}^{\dagger}u_{11}^{\dagger}u_{11}\hat{U}_{\beta}|\alpha\rangle|\tilde{\gamma}\rangle.
\label{probability-ground-state-thermal-2}
\end{equation}
Moreover, using Eqs.~(\ref{unitary-evolution-2}) and (\ref{unitary-evolution-3}),
we can rewrite the Hermitian operator $u_{11}^{\dagger}u_{11}$ as
\begin{equation}
u_{11}^{\dagger}u_{11}
=
\cos^{2}(|\kappa|t\sqrt{a^{\dagger}a+c})
+
c
\frac{\sin^{2}(|\kappa|t\sqrt{a^{\dagger}a+c})}{a^{\dagger}a+c},
\label{explicit-form-u11+u11-1}
\end{equation}
where $c$ is defined in Eq.~(\ref{definition-constant-c}).

To compute $P_{g}(\beta;t)$ approximately,
we formulate the perturbation theory as follows.
At first, we consider an arbitrary function $g(x)$,
which can be represented as a Taylor series about $x=0$,
\begin{equation}
g(x)
=
\sum_{n=0}^{\infty}g^{(n)}x^{n},
\label{g-function-Taylor-series-i}
\end{equation}
where $-\infty<x<+\infty$ and $g^{(n)}$ is an arbitrary complex number for all $n$.
[In Eq.~(\ref{g-function-Taylor-series-i}),
$g^{(n)}$ is equal to
$(d^{n}/dx^{n})g(x)|_{x=0}$.
This notation is slightly different from $f^{(m)}(n)$ given by Eq.~(\ref{Taylor-series-01}).]
Next, we evaluate $\hat{U}_{\beta}^{\dagger}g(a^{\dagger}a+c)\hat{U}_{\beta}$
by the second-order perturbation theory at low temperature.
We assume that the temperature $T$ is quite low,
so that $\beta\epsilon=\epsilon(k_{\mbox{\scriptsize B}}T)^{-1}\gg 1$.
Then, from $\beta\epsilon\gg 1$ and Eq.~(\ref{definition-theta-beta-1}),
we understand $0<\theta(\beta)\ll 1$.
Decomposing $\tanh\theta(\beta)$
into the power series in the small parameter $\theta(\beta)$,
\begin{eqnarray}
\tanh\theta(\beta)
&=&
e^{-\beta\epsilon/2} \nonumber \\
&=&
\theta(\beta)-(1/3)\theta(\beta)^{3}+O(\theta^{5}),
\end{eqnarray}
we obtain
$\theta(\beta)\simeq e^{-\beta\epsilon/2}$
with neglecting terms of order $O(\theta^{3})$.
From now on,
we consider the second-order perturbation theory with the parameter $\theta(\beta)$.

From the above discussions,
we can compute $\hat{U}_{\beta}^{\dagger}g(a^{\dagger}a+c)\hat{U}_{\beta}$ approximately
as follows:
\begin{eqnarray}
&&
\hat{U}_{\beta}^{\dagger}g(a^{\dagger}a+c)\hat{U}_{\beta} \nonumber \\
&=&
g(a^{\dagger}a+c)
+
\theta(\beta)
\sum_{n=0}^{\infty}g^{(n)}
[a^{\dagger}\tilde{a}^{\dagger}-a\tilde{a},(a^{\dagger}a+c)^{n}] \nonumber \\
&&\quad
+
(1/2)\theta(\beta)^{2}
\sum_{n=0}^{\infty}g^{(n)}
[a^{\dagger}\tilde{a}^{\dagger}-a\tilde{a},[a^{\dagger}\tilde{a}^{\dagger}-a\tilde{a},(a^{\dagger}a+c)^{n}]] \nonumber \\
&&\quad
+
O(\theta^{3}).
\label{UgU-second-order-perturbation-i}
\end{eqnarray}
(In the above derivation,
we use the Baker-Hausdorff theorem~\cite{Louisell1973}.)
In the following subsections,
we calculate the first and second-order perturbation corrections, respectively.

\subsection{\label{subsection-first-order-correction}
The first-order correction}

In this subsection, we calculate the first-order correction given by Eq.~(\ref{UgU-second-order-perturbation-i}).
To obtain the commutation relation of
$(a^{\dagger}\tilde{a}^{\dagger}-a\tilde{a})$
and
$(a^{\dagger}a+c)^{n}$,
we use the following notation:
\begin{equation}
\hat{A}
=
a^{\dagger}\tilde{a}^{\dagger}-a\tilde{a},
\quad
\hat{B}
=
a^{\dagger}a+c,
\quad
\hat{C}
=
a^{\dagger}\tilde{a}^{\dagger}+a\tilde{a}.
\end{equation}
Then, we obtain the following relations:
\begin{equation}
[\hat{A},\hat{B}]=-\hat{C},
\quad
[\hat{C},\hat{B}]=-\hat{A}.
\label{commutation-relations-ABC}
\end{equation}
Using Eq.~(\ref{commutation-relations-ABC}), we obtain
\begin{eqnarray}
[\hat{A},\hat{B}]
&=&
-\hat{C}, \nonumber \\
{[}\hat{A},\hat{B}^{2}{]}
&=&
\hat{A}-2\hat{B}\hat{C}, \nonumber \\
{[}\hat{A},\hat{B}^{3}{]}
&=&
-\hat{C}+3\hat{B}\hat{A}-3\hat{B}^{2}\hat{C}, \nonumber \\
{[}\hat{A},\hat{B}^{4}{]}
&=&
\hat{A}-4\hat{B}\hat{C}+6\hat{B}^{2}\hat{A}-4\hat{B}^{3}\hat{C},
\label{commutation-relations-A-Bn-1}
\end{eqnarray}
and so on.

Here, we introduce the following operators:
\begin{equation}
\hat{\mu}=a^{\dagger}\tilde{a}^{\dagger},
\quad
\hat{\nu}=a\tilde{a}.
\label{definition-operators-mu-nu-1}
\end{equation}
Because
\begin{equation}
\hat{A}=\hat{\mu}-\hat{\nu},
\quad
\hat{C}=\hat{\mu}+\hat{\nu},
\label{definition-operators-mu-nu-2}
\end{equation}
we can rewrite Eq.~(\ref{commutation-relations-A-Bn-1}) as follows:
\begin{eqnarray}
[\hat{A},\hat{B}]
&=&
(\hat{B}-1)\hat{\mu}-(1+\hat{B})\hat{\nu}-\hat{B}\hat{A}, \nonumber \\
{[}\hat{A},\hat{B}^{2}{]}
&=&
(\hat{B}-1)^{2}\hat{\mu}-(1+\hat{B})^{2}\hat{\nu}-\hat{B}^{2}\hat{A}, \nonumber \\
{[}\hat{A},\hat{B}^{3}{]}
&=&
(\hat{B}-1)^{3}\hat{\mu}-(1+\hat{B})^{3}\hat{\nu}-\hat{B}^{3}\hat{A}, \nonumber \\
{[}\hat{A},\hat{B}^{4}{]}
&=&
(\hat{B}-1)^{4}\hat{\mu}-(1+\hat{B})^{4}\hat{\nu}-\hat{B}^{4}\hat{A}.
\label{commutation-relations-A-Bn-2}
\end{eqnarray}
Thus, we obtain
\begin{equation}
[\hat{A},\hat{B}^{n}]
=
(\hat{B}-1)^{n}\hat{\mu}-(\hat{B}+1)^{n}\hat{\nu}-\hat{B}^{n}\hat{A}
\quad\mbox{for $n=1,2,3,...$}.
\label{commutation-relations-A-Bn-3}
\end{equation}
From Eqs.~(\ref{commutation-relations-A-Bn-2})
and (\ref{commutation-relations-A-Bn-3}),
we obtain the first-order correction as
\begin{eqnarray}
&&
\theta(\beta)
\sum_{n=0}^{\infty}g^{(n)}
[a^{\dagger}\tilde{a}^{\dagger}-a\tilde{a},(a^{\dagger}a+c)^{n}]  \nonumber \\
&=&
\theta(\beta)
[
g(\hat{B}-1)\hat{\mu}
-
g(\hat{B}+1)\hat{\nu}
-
g(\hat{B})\hat{A}
].
\label{g-first-order-perturbation-ii}
\end{eqnarray}

\subsection{\label{subsection-second-order-correction}
The second-order correction}

In this subsection, we calculate the second-order correction given by Eq.~(\ref{UgU-second-order-perturbation-i}).
From Eq.~(\ref{commutation-relations-A-Bn-3}),
we obtain
\begin{equation}
[\hat{A},[\hat{A},\hat{B}^{n}]] \nonumber \\
=
\hat{R}_{n}+\hat{S}_{n}
\quad\mbox{for $n=1,2,3,...$},
\end{equation}
where
\begin{eqnarray}
\hat{R}_{n}
&=&
[\hat{A},(\hat{B}-1)^{n}]\hat{\mu}
-
[\hat{A},(\hat{B}+1)^{n}]\hat{\nu}
-
[\hat{A},\hat{B}^{n}]\hat{A}, \nonumber \\
\hat{S}_{n}
&=&
(\hat{B}-1)^{n}[\hat{A},\hat{\mu}]
+
(\hat{B}+1)^{n}[\hat{A},\hat{\nu}].
\label{defintion-Rn-Sn-i}
\end{eqnarray}
Thus, we can divide the second-order correction up into the following two parts:
\begin{equation}
\frac{1}{2}\theta(\beta)^{2}\sum_{n=0}^{\infty}g^{(n)}\hat{R}_{n}
+
\frac{1}{2}\theta(\beta)^{2}\sum_{n=0}^{\infty}g^{(n)}\hat{S}_{n}.
\label{second-order-correction-i}
\end{equation}

First, we evaluate the part including $\{\hat{S}_{n}\}$
in Eq.~(\ref{second-order-correction-i}).
Because of Eqs.~(\ref{definition-operators-mu-nu-1})
and (\ref{definition-operators-mu-nu-2}),
we obtain
\begin{equation}
[\hat{A},\hat{\mu}]=[\hat{A},\hat{\nu}]=-\hat{D},
\end{equation}
where
\begin{equation}
\hat{D}=a^{\dagger}a+\tilde{a}^{\dagger}\tilde{a}+1.
\end{equation}
[We note that the above $\hat{D}$ is different from $D$ defined in Eq.~(\ref{unitary-evolution-3}).]
Thus, we can write down $\hat{S}_{n}$ as
\begin{equation}
\hat{S}_{n}
=
-
\Bigl(
(\hat{B}-1)^{n}+(\hat{B}+1)^{n}
\Bigr)
\hat{D}.
\end{equation}
Hence, we can rewrite the term that includes $\{\hat{S}_{n}\}$
in Eq.~(\ref{second-order-correction-i}) as
\begin{equation}
-\frac{1}{2}\theta(\beta)^{2}
\Bigr(
g(\hat{B}-1)+g(\hat{B}+1)
\Bigr)
\hat{D}.
\label{second-order-correction-S-part-i}
\end{equation}

Second, we evaluate the part including $\{\hat{R}_{n}\}$
in Eq.~(\ref{second-order-correction-i}).
From Eq.~(\ref{defintion-Rn-Sn-i}),
we can confirm $\hat{R}_{0}=\hat{R}_{1}=0$ at ease.
After slightly tough calculations,
we obtain
\begin{eqnarray}
\hat{R}_{2}
&=&
-2[\hat{A},\hat{B}]\hat{C}, \nonumber \\
\hat{R}_{3}
&=&
-3[\hat{A},\hat{B}^{2}]\hat{C}
+3[\hat{A},\hat{B}]\hat{A}, \nonumber \\
\hat{R}_{4}
&=&
-4[\hat{A},\hat{B}^{3}]\hat{C}
+6[\hat{A},\hat{B}^{2}]\hat{A}
-4[\hat{A},\hat{B}]\hat{C},
\end{eqnarray}
and so on.
Thus, we can write down the part that includes $\{\hat{R}_{n}\}$
in Eq.~(\ref{second-order-correction-i}) as
\begin{eqnarray}
&&
\frac{1}{2}\theta(\beta)^{2}
\Bigr(
-
\sum_{n=0}^{\infty}
(n+2)g^{(n+2)}[\hat{A},\hat{B}^{n+1}]\hat{C} \nonumber \\
&&\quad
+
\frac{1}{2}\sum_{n=0}^{\infty}
(n+2)(n+3)g^{(n+3)}[\hat{A},\hat{B}^{n+1}]\hat{A} \nonumber \\
&&\quad
-
\frac{1}{6}\sum_{n=0}^{\infty}
(n+2)(n+3)(n+4)g^{(n+4)}[\hat{A},\hat{B}^{n+1}]\hat{C}
+
...
\Bigr) \nonumber \\
&=&
\frac{1}{2}\theta(\beta)^{2}
\Bigr(
-\hat{F}_{1}\hat{C}
+\frac{1}{2}\hat{F}_{2}\hat{A}
-\frac{1}{3!}\hat{F}_{3}\hat{C}
+...
\Bigr) \nonumber \\
&=&
\frac{1}{2}\theta(\beta)^{2}
\Bigr(
\sum_{n=0}^{\infty}
\frac{(-1)^{n}}{n!}\hat{F}_{n}\hat{\mu}
-
\sum_{n=0}^{\infty}
\frac{1}{n!}\hat{F}_{n}\hat{\nu}
\Bigr) \nonumber \\
&&\quad
-
\frac{1}{2}\theta(\beta)^{2}
\Bigr(
g(\hat{B}-1)\hat{\mu}-g(\hat{B}+1)\hat{\nu}-g(\hat{B})\hat{A}
\Bigr)\hat{A},
\label{second-order-correction-R-part-i}
\end{eqnarray}
where
\begin{equation}
\hat{F}_{n}
=
\frac{d^{n}}{dx^{n}}g(x)\bigg|_{x=\hat{B}-1}\hat{\mu}
-
\frac{d^{n}}{dx^{n}}g(x)\bigg|_{x=\hat{B}+1}\hat{\nu}
-
\frac{d^{n}}{dx^{n}}g(x)\bigg|_{x=\hat{B}}\hat{A}.
\label{definition-operator-Gn}
\end{equation}
[In the above derivation,
we use Eq.~(\ref{commutation-relations-A-Bn-3}) in an effective manner.
The form of $\hat{F}_{n}$ in Eq.~(\ref{definition-operator-Gn})
reflects Eq.~(\ref{commutation-relations-A-Bn-3}).]

Here, we introduce the following operators:
\begin{equation}
e^{\pm d/dx}g(x)
=
\sum_{n=0}^{\infty}
\frac{(-1)^{n}}{n!}
\frac{d^{n}}{dx^{n}}g(x).
\end{equation}
Using the symbols $e^{\pm d/dx}$,
we can rewrite the first term
of Eq.~(\ref{second-order-correction-R-part-i}) as
\begin{eqnarray}
&&
\frac{1}{2}\theta(\beta)^{2}
\Bigr(
[
e^{-d/dx}g(\hat{B}-1)\hat{\mu}
-
e^{-d/dx}g(\hat{B}+1)\hat{\nu}
-
e^{-d/dx}g(\hat{B})\hat{A}
]\hat{\mu} \nonumber \\
&&\quad
-
[
e^{d/dx}g(\hat{B}-1)\hat{\mu}
-
e^{d/dx}g(\hat{B}+1)\hat{\nu}
-
e^{d/dx}g(\hat{B})\hat{A}
]\hat{\nu}
\Bigr).
\label{second-order-correction-R-part-ii}
\end{eqnarray}
Then, we apply the following technique
to Eq.~(\ref{second-order-correction-R-part-ii}):
\begin{equation}
e^{\pm d/dx}g(\hat{X})
=
\sum_{n=0}^{\infty}
\frac{(-1)^{n}}{n!}
\frac{d^{n}}{dx^{n}}g(x)\bigg|_{x=\hat{X}}
=
g(\hat{X}\pm 1),
\end{equation}
where $\hat{X}$ is an arbitrary operator.
Thus, we can rewrite Eq.~(\ref{second-order-correction-R-part-ii})
as
\begin{eqnarray}
&&
\frac{1}{2}\theta(\beta)^{2}
\Bigr(
[
g(\hat{B}-2)\hat{\mu}
-
g(\hat{B})\hat{\nu}
-
g(\hat{B}-1)\hat{A}
]\hat{\mu} \nonumber \\
&&\quad
-
[
g(\hat{B})\hat{\mu}
-
g(\hat{B}+2)\hat{\nu}
-
g(\hat{B}+1)\hat{A}
]\hat{\nu}
\Bigr).
\label{second-order-correction-R-part-iii}
\end{eqnarray}

Finally, from Eqs.~(\ref{second-order-correction-i}),
(\ref{second-order-correction-S-part-i}),
(\ref{second-order-correction-R-part-i}) and
(\ref{second-order-correction-R-part-iii}),
we obtain the second-order correction as
\begin{eqnarray}
&&
(1/2)\theta(\beta)^{2}
\sum_{n=0}^{\infty}g^{(n)}
[a^{\dagger}\tilde{a}^{\dagger}-a\tilde{a},
[a^{\dagger}\tilde{a}^{\dagger}-a\tilde{a},(a^{\dagger}a+c)^{n}]] \nonumber \\
&=&
(1/2)\theta(\beta)^{2}
[
g(\hat{B}-2)\hat{\mu}^{2}
+
g(\hat{B}-1)(-\hat{D}-\hat{\mu}\hat{A}-\hat{A}\hat{\mu}) \nonumber \\
&&\quad
+
g(\hat{B})(\hat{A}^{2}-\hat{\nu}\hat{\mu}-\hat{\mu}\hat{\nu})
+
g(\hat{B}+1)(-\hat{D}+\hat{\nu}\hat{A}+\hat{A}\hat{\nu})
+
g(\hat{B}+2)\hat{\nu}^{2}].
\label{second-order-correction-ii}
\end{eqnarray}

\subsection{\label{subsection-atomic-population-inversion-second-order-perturbation}
The atomic population inversion up to the second-order correction}

In this subsection,
we compute $P_{g}(\beta;t)$ defined
in Eq.~(\ref{probability-ground-state-thermal-2})
up to the second-order perturbation.
According to Eq.~(\ref{explicit-form-u11+u11-1}),
we define $g(x)$ as follows:
\begin{equation}
g(x+c)
=
\cos^{2}(|\kappa|t\sqrt{x+c})
+
\frac{c}{x+c}
\sin^{2}(|\kappa|t\sqrt{x+c}).
\end{equation}
From Eqs.~(\ref{probability-ground-state-thermal-2}),
(\ref{UgU-second-order-perturbation-i}),
(\ref{g-first-order-perturbation-ii})
and (\ref{second-order-correction-ii}),
we obtain $P_{g}(\beta;t)$ as
\begin{equation}
P_{g}(\beta;t)
=
P_{g}(t)
+
\theta(\beta)P^{(1)}_{g}(\beta;t)
+
(1/2)\theta(\beta)^{2}P^{(2)}_{g}(\beta;t)
+
O(\theta^{3}),
\end{equation}
where
\begin{eqnarray}
P^{(1)}_{g}(\beta;t)
&=&
\langle\alpha|\langle\tilde{\gamma}|
[g(\hat{B}-1)\hat{\mu}-g(\hat{B}+1)\hat{\nu}-g(\hat{B})\hat{A}]
|\alpha\rangle|\tilde{\gamma}\rangle, \nonumber \\
P^{(2)}_{g}(\beta;t)
&=&
\langle\alpha|\langle\tilde{\gamma}|
[g(\hat{B}-2)\hat{\mu}^{2}
+
g(\hat{B}-1)(-\hat{D}-\hat{\mu}\hat{A}-\hat{A}\hat{\mu}) \nonumber \\
&&\quad
+
g(\hat{B})(\hat{A}^{2}-\hat{\nu}\hat{\mu}-\hat{\mu}\hat{\nu})
+
g(\hat{B}+1)(-\hat{D}+\hat{\nu}\hat{A}+\hat{A}\hat{\nu}) \nonumber \\
&&\quad
+
g(\hat{B}+2)\hat{\nu}^{2}]|\alpha\rangle|\tilde{\gamma}\rangle.
\end{eqnarray}

First, we compute $P^{(1)}_{g}(\beta;t)$.
Remembering that $|\alpha\rangle$ and $|\tilde{\gamma}\rangle$ are coherent states
on ${\cal H}$ and $\tilde{{\cal H}}$ respectively,
we can write down $P^{(1)}_{g}(\beta;t)$ as follows:
\begin{eqnarray}
P^{(1)}_{g}(\beta;t)
&=&
\langle\alpha|\langle\tilde{\gamma}|
\sum_{n=0}^{\infty}\sum_{m=0}^{\infty}
\exp(-\frac{\alpha^{2}+\tilde{\gamma}^{2}}{2})
\frac{\alpha^{n}}{\sqrt{n!}}
\frac{\tilde{\gamma}^{m}}{\sqrt{m!}} \nonumber \\
&&\quad\times
\Bigl(
[g(a^{\dagger}a+c-1)-g(a^{\dagger}a+c)]a^{\dagger}\tilde{a}^{\dagger} \nonumber \\
&&\quad
-
[g(a^{\dagger}a+c+1)-g(a^{\dagger}a+c)]a\tilde{a}
\Bigr)
|n\rangle|\tilde{m}\rangle \nonumber \\
&=&
\exp[-(\alpha^{2}+\tilde{\gamma}^{2})] \nonumber \\
&&\quad\times
\Bigl(
\sum_{n=0}^{\infty}\sum_{m=0}^{\infty}
\frac{\alpha^{2n+1}}{n!}
\frac{\tilde{\gamma}^{2m+1}}{m!}
[g(n+c)-g(n+c+1)] \nonumber \\
&&\quad
-
\sum_{n=1}^{\infty}\sum_{m=1}^{\infty}
\frac{\alpha^{2n-1}}{(n-1)!}
\frac{\tilde{\gamma}^{2m-1}}{(m-1)!}
[g(n+c)-g(n+c-1)]
\Bigr) \nonumber \\
&=&
2\alpha\tilde{\gamma}
\exp(-\alpha^{2})
\sum_{n=0}^{\infty}
\frac{\alpha^{2n}}{n!}
[g(n+c)-g(n+c+1)] \nonumber \\
&=&
2\alpha\tilde{\gamma}
[P_{g}(t)-Q^{(1)}_{g}(t)],
\label{P1g-beta-t-final-form}
\end{eqnarray}
where
\begin{eqnarray}
Q^{(l)}_{g}(t)
&=&
\exp(-\alpha^{2})
\sum_{n=0}^{\infty}\frac{\alpha^{2n}}{n!}
[\cos^{2}(\sqrt{c+n+l}|\kappa|t)
+
\frac{c}{c+n+l}
\sin^{2}(\sqrt{c+n+l}|\kappa|t)] \nonumber \\
&=&
\exp(-\alpha^{2})
|\alpha|^{-2(c+l)}
\sum_{n=0}^{\infty}\frac{|\alpha|^{2(n+c+l)}}{n!}
[\cos^{2}(\sqrt{c+n+l}|\kappa|t) \nonumber \\
&&\quad
+
\frac{c}{c+n+l}
\sin^{2}(\sqrt{c+n+l}|\kappa|t)] \nonumber \\
&=&
\exp(-\alpha^{2})
\Bigl[
\frac{1}{2}
[\cos^{2}(|\kappa|t\sqrt{c+l})+\frac{c}{c+l}\sin^{2}(|\kappa|t\sqrt{c+l})] \nonumber \\
&&\quad
+
I^{(l)}_{1}(t)-2I^{(l)}_{2}(t)
\Bigr]
\quad
\mbox{for $l=0,1,2,...$},
\label{Qlgt-integral-form}
\end{eqnarray}
and
$I^{(l)}_{1}(t)$ and $I^{(l)}_{2}(t)$ are defined in Eq.~(\ref{definition-Il1t-Il2t-i}).
We note that $Q^{(0)}_{g}(t)=P_{g}(t)$.

Second, we compute $P^{(2)}_{g}(\beta;t)$.
Before obtaining an explicit form of $P^{(2)}_{g}(\beta;t)$,
we prepare the following relations:
\begin{eqnarray}
&&
\hat{\mu}^{2}|n\rangle|\tilde{m}\rangle \nonumber \\
&=&
\sqrt{(n+1)(n+2)(m+1)(m+2)}|n+2\rangle|\widetilde{m+2}\rangle, \nonumber \\
&&
(-\hat{D}-\hat{\mu}\hat{A}-\hat{A}\hat{\mu})|n\rangle|\tilde{m}\rangle  \nonumber \\
&=&
2nm|n\rangle|\tilde{m}\rangle
-
2\sqrt{(n+1)(n+2)(m+1)(m+2)}|n+2\rangle|\widetilde{m+2}\rangle, \nonumber \\
&&
(\hat{A}^{2}-\hat{\nu}\hat{\mu}-\hat{\mu}\hat{\nu})|n\rangle|\tilde{m}\rangle  \nonumber \\
&=&
\sqrt{(n+1)(n+2)(m+1)(m+2)}|n+2\rangle|\widetilde{m+2}\rangle \nonumber \\
&&\quad
-
2(2nm+n+m+1)|n\rangle|\tilde{m}\rangle
+
\sqrt{n(n-1)m(m-1)}|n-2\rangle|\widetilde{m-2}\rangle, \nonumber \\
&&
(-\hat{D}+\hat{\nu}\hat{A}+\hat{A}\hat{\nu})|n\rangle|\tilde{m}\rangle  \nonumber \\
&=&
2nm|n\rangle|\tilde{m}\rangle
-
2\sqrt{n(n-1)m(m-1)}|n-2\rangle|\widetilde{m-2}\rangle, \nonumber \\
&&
\hat{\nu}^{2}|n\rangle|\tilde{m}\rangle  \nonumber \\
&=&
\sqrt{n(n-1)m(m-1)}|n-2\rangle|\widetilde{m-2}\rangle
\quad
\mbox{for $n,m\in\{0,1,2,...\}$}.
\label{relations-number-states-H-tildeH}
\end{eqnarray}
Using Eq.~(\ref{relations-number-states-H-tildeH}),
we can write down $P^{(2)}_{g}(\beta;t)$ as
\begin{eqnarray}
P^{(2)}_{g}(\beta;t)
&=&
\frac{1}{2}\theta(\beta)^{2}
\exp(-\frac{\alpha^{2}+\tilde{\gamma}^{2}}{2})
\langle\alpha|\langle\tilde{\gamma}|
\sum_{n=0}^{\infty}\sum_{m=0}^{\infty}
\frac{\alpha^{n}}{\sqrt{n!}}\frac{\tilde{\gamma}^{m}}{\sqrt{m!}} \nonumber \\
&&\quad
\times
\Bigl(
[g(n+c)-2g(n+c+1)+g(n+c+2)] \nonumber \\
&&\quad
\times
\sqrt{(n+1)(n+2)(m+1)(m+2)}|n+2\rangle|\widetilde{m+2}\rangle \nonumber \\
&&\quad
+
[2nmg(n+c-1)-2(2nm+n+m+1)g(n+c) \nonumber \\
&&\quad
+2nmg(n+c+1)]|n\rangle|\tilde{m}\rangle \nonumber \\
&&\quad
+
[g(n+c-2)-2g(n+c-1)+g(n+c)] \nonumber \\
&&\quad
\times
\sqrt{n(n-1)m(m-1)}|n-2\rangle|\widetilde{m-2}\rangle
\Bigr) \nonumber \\
&=&
\frac{1}{2}\theta(\beta)^{2}\exp[-(\alpha^{2}+\tilde{\gamma}^{2})]
\sum_{n=0}^{\infty}\sum_{m=0}^{\infty} \nonumber \\
&&\quad
\times
\Bigl(
\frac{\alpha^{2n+2}}{n!}\frac{\tilde{\gamma}^{2m+2}}{m!}
[g(n+c)-2g(n+c+1)+g(n+c+2)] \nonumber \\
&&\quad
+
\frac{\alpha^{2n}}{n!}\frac{\tilde{\gamma}^{2m}}{m!}
2[nmg(n+c+1)-(2nm+n+m+1)g(n+c) \nonumber \\
&&\quad
+nmg(n+c-1)] \nonumber \\
&&\quad
+
\frac{\alpha^{2n-2}}{(n-2)!}\frac{\tilde{\gamma}^{2m-2}}{(m-2)!}
[g(n+c)-2g(n+c-1)+g(n+c-2)]
\Bigr) \nonumber \\
&=&
\frac{1}{2}\theta(\beta)^{2}\exp(-\alpha^{2})
\sum_{n=0}^{\infty}
\frac{\alpha^{2n}}{n!} \nonumber \\
&&\quad
\times
\Bigl(
4\tilde{\gamma}^{2}\alpha^{2}
[g(n+c)-2g(n+c+1)+g(n+c+2)] \nonumber \\
&&\quad
-
2\alpha^{2}g(n+c+1)
-
2\tilde{\gamma}^{2}g(n+c)
-
2g(n+c)
\Bigr) \nonumber \\
&=&
\frac{1}{2}\theta(\beta)^{2}
[
2(2\alpha^{2}\tilde{\gamma}^{2}-\tilde{\gamma}^{2}-1)P_{g}(t)
-
2\alpha^{2}(4\tilde{\gamma}^{2}+1)Q^{(1)}_{g}(t) \nonumber \\
&&\quad
+
4\alpha^{2}\tilde{\gamma}^{2}Q^{(2)}_{g}(t)].
\label{P2g-beta-t-final-form}
\end{eqnarray}

\section{\label{section-Properties-integrals}
Properties of the integrals that form the atomic population inversion}

In this section, numerically evaluating the integrals that form the atomic population inversion,
we examine their physical meanings.

\begin{figure}
\begin{center}
\includegraphics[scale=1.0]{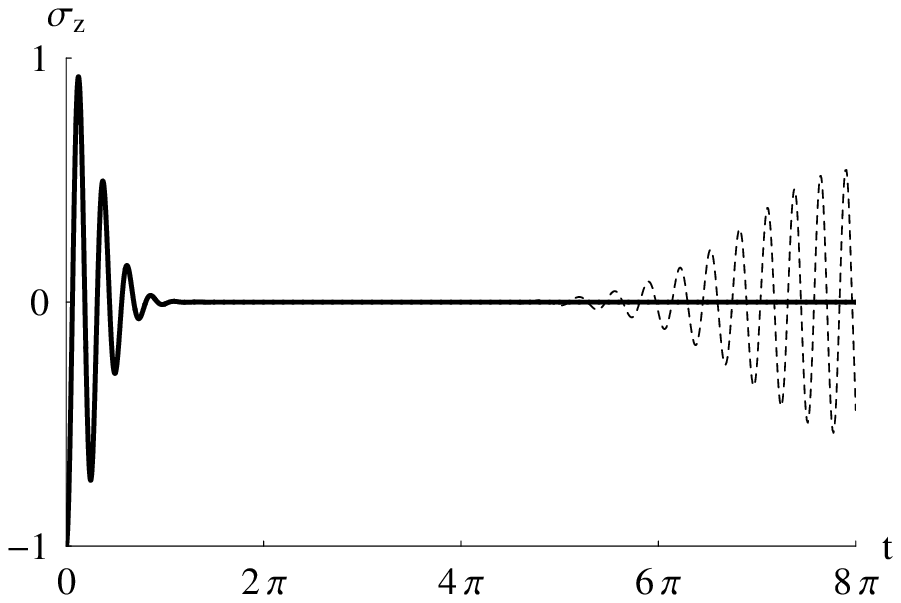}
\end{center}
\caption{The graphs of $\langle\sigma_{z}(t)\rangle$
given by Eq.~(\ref{atomic-population-inversion-equation01})
and $J_{1}(t)$ given by Eq.~(\ref{definition-first-and-second-integral00})
for $t\in[0,8\pi]$ and $\alpha=4$.
(We consider the resonant case,
so that we assume $\Delta\omega=0$ and $\kappa=1$.)
A thick solid curve represents $J_{1}(t)$
and a thin dashed curve represents $\langle\sigma_{z}(t)\rangle$.
In the numerical calculation of $\langle\sigma_{z}(t)\rangle$
given by Eq.~(\ref{atomic-population-inversion-equation01}),
the summation of the index $n$ is carried out up to $n=100$.
In the numerical calculation of $J_{1}(t)$
given by Eq.~(\ref{definition-first-and-second-integral00}),
the integral $\int_{0}^{\infty}dx$ is replaced with $\int_{0}^{100}dx$.
The interval of the numerical integration $x\in[0,100]$ is divided into $10^{6}$ steps
($\Delta x=1.0\times 10^{-4}$) and we apply Simpson's rule.
To obtain the variation of $J_{1}(t)$ against the time $t$,
we divide the interval $t\in[0,8\pi]$ into $4000$ steps
($\Delta t=2\pi\times 10^{-3}$)
and we estimate $J_{1}(t)$ at each time step.
Looking at this figure, we notice the following facts.
In the graph of $J_{1}(t)$, we can observe only the initial collapse
and we cannot observe the revival of the Rabi oscillations.
[$\langle\sigma_{z}(t)\rangle$ starts to show the revival of the Rabi oscillations
around at $t=5\pi$.
By contrast, $J_{1}(t)$ is nearly equal to zero after $t=\pi$.]}
\label{Figure02}
\end{figure}

\begin{figure}
\begin{center}
\includegraphics[scale=1.0]{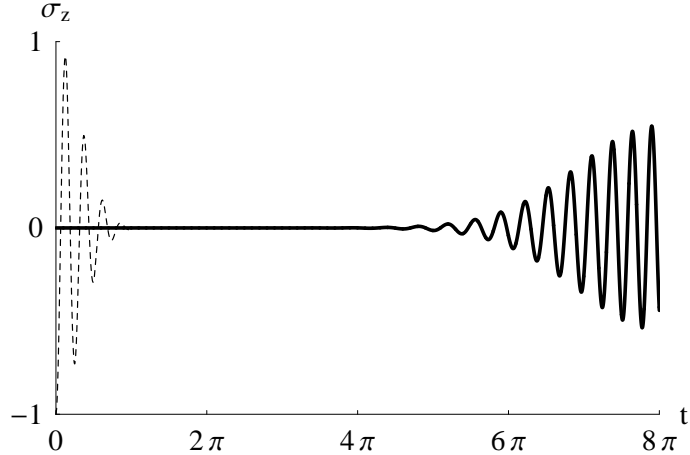}
\end{center}
\caption{The graphs of $\langle\sigma_{z}(t)\rangle$
given by Eq.~(\ref{atomic-population-inversion-equation01})
and $J_{2}(t)$ given by Eq.~(\ref{definition-first-and-second-integral00})
for $t\in[0,8\pi]$ and $\alpha=4$.
(We consider the resonant case,
so that we assume $\Delta\omega=0$ and $\kappa=1$.)
A thick solid curve represents $J_{2}(t)$
and a thin dashed curve represents $\langle\sigma_{z}(t)\rangle$.
In the numerical calculation of $\langle\sigma_{z}(t)\rangle$
given by Eq.~(\ref{atomic-population-inversion-equation01}),
the summation of the index $n$ is carried out up to $n=100$.
In the numerical calculation of $J_{2}(t)$
given by Eq.~(\ref{definition-first-and-second-integral00}),
the integral $\int_{0}^{\infty}dy$ is replaced with $\int_{0}^{100}dy$.
The interval of the numerical integration $y\in[0,100]$ is divided into $10^{6}$ steps
($\Delta y=1.0\times 10^{-4}$) and we apply Bode's rule.
To obtain the variation of $J_{2}(t)$ against the time $t$,
we divide the interval $t\in[0,8\pi]$ into $4000$ steps
($\Delta t=2\pi\times 10^{-3}$)
and we estimate $J_{2}(t)$ at each time step.
Looking at this figure, we notice the following facts.
In the graph of $J_{2}(t)$, we can observe only the revival
and we cannot observe the initial collapse of the Rabi oscillations.
[$\langle\sigma_{z}(t)\rangle$ shows the initial collapse from $t=0$
until around $t=\pi$.
By contrast, $J_{2}(t)$ is nearly equal to zero from $t=0$ to $t=4\pi$.]}
\label{Figure03}
\end{figure}

First,
we consider the case where the electromagnetic field is resonant with the atom
at zero-temperature,
so that we put $\Delta\omega=0$ and $\kappa=1$.
We show graphs of $J_{1}(t)$ defined in Eq.~(\ref{definition-first-and-second-integral00})
and $\langle\sigma_{z}(t)\rangle$ given by
Eq.~(\ref{atomic-population-inversion-equation01})
in Fig.~\ref{Figure02}.
We show graphs of $J_{2}(t)$ defined in Eq.~(\ref{definition-first-and-second-integral00})
and $\langle\sigma_{z}(t)\rangle$ given by
Eq.~(\ref{atomic-population-inversion-equation01})
in Fig.~\ref{Figure03}.
In these estimations, we put $\alpha=4$.
Thus, we can neglect the first term of the right-hand side of
Eq.~(\ref{atomic-population-inversion-equation03})
[$-(1/2)\exp(-\alpha^{2})=-5.63\times 10^{-8}$].

Looking at Fig.~\ref{Figure02},
we can conclude that $J_{1}(t)$
only represents the initial collapse
(the semi-classical limit) and it does not represent the revival (the quantum correction)
of the Rabi oscillations.
By contrast, looking at Fig.~\ref{Figure03},
we can conclude that $J_{2}(t)$
only represents the revival
and it does not represent the initial collapse
of the Rabi oscillations.
In our numerical calculations,
we obtain
$|\langle\sigma_{z}(t)\rangle-J_{1}(t)|\leq 8.0847\times10^{-4}$
for $t\in[0,4\pi]$
and
$|\langle\sigma_{z}(t)\rangle-J_{2}(t)|\leq 5.6377\times10^{-8}$
for $t\in[4\pi,8\pi]$.

Here, we pay attention to the representation of $J_{1}(t)$
given in Eq.~(\ref{definition-first-and-second-integral00}).
If we let $t$ be a large value ($t\gg 1$),
the trigonometric function $\cos(2\sqrt{x}t)$ included in the integrand
oscillates intensely and rapidly for the small variation of $x$.
Thus, the integral $\int_{0}^{\infty}dx$ of $J_{1}(t)$
converges on zero for $t\gg 1$.
Therefore, we can expect that $J_{1}(t)$ never causes the revival of the Rabi oscillations.
Hence, we can expect that $J_{2}(t)$ lets the revival of the Rabi oscillations
happen in the range of $t\gg 1$.

\begin{figure}
\begin{center}
\includegraphics[scale=1.0]{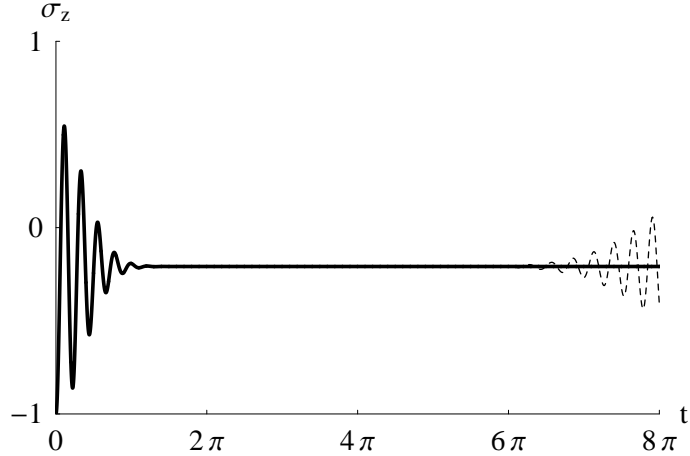}
\end{center}
\caption{The graphs of $\langle\sigma_{z}(t)\rangle$
given by Eqs.~(\ref{probability-ground-state-01})
and (\ref{atomic-population-inversion-equation00})
and
$[1-2\exp(-\alpha^{2})I^{(0)}_{1}(t)]$
given by Eq.~(\ref{definition-Il1t-Il2t-i})
for $t\in[0,8\pi]$, $\alpha=4$, $\Delta\omega=4$, $\kappa=1$ and $c=4$.
(We consider the off-resonant case.)
A thick solid curve represents $[1-2\exp(-\alpha^{2})I^{(0)}_{1}(t)]$
and a thin dashed curve represents $\langle\sigma_{z}(t)\rangle$.
In the numerical calculation of $\langle\sigma_{z}(t)\rangle$
given by Eqs.~(\ref{probability-ground-state-01})
and (\ref{atomic-population-inversion-equation00}),
the summation of the index $n$ is carried out up to $n=100$.
In the numerical calculation of $[1-2\exp(-\alpha^{2})I^{(0)}_{1}(t)]$
given by Eq.~(\ref{definition-Il1t-Il2t-i}),
the integral $\int_{0}^{\infty}dx$ is replaced with $\int_{0}^{100}dx$.
The interval of the numerical integration $x\in[0,100]$ is divided into $10^{6}$ steps
($\Delta x=1.0\times 10^{-4}$) and we apply Simpson's rule.
To obtain the variation of $[1-2\exp(-\alpha^{2})I^{(0)}_{1}(t)]$ against the time $t$,
we divide the interval $t\in[0,8\pi]$ into $4000$ steps
($\Delta t=2\pi\times 10^{-3}$)
and we estimate $[1-2\exp(-\alpha^{2})I^{(0)}_{1}(t)]$ at each time step.
Looking at this figure, we notice the following facts.
In the graph of $[1-2\exp(-\alpha^{2})I^{(0)}_{1}(t)]$, we can observe only the initial collapse
and we cannot observe the revival of the Rabi oscillations.
[$\langle\sigma_{z}(t)\rangle$ starts to show the revival of the Rabi oscillations
around at $t=6\pi$.
By contrast, $[1-2\exp(-\alpha^{2})I^{(0)}_{1}(t)]$ is nearly equal to $\mbox{Const.}$
given by Eq.(\ref{definition-Const}) after $t=\pi$.]}
\label{Figure04}
\end{figure}

\begin{figure}
\begin{center}
\includegraphics[scale=1.0]{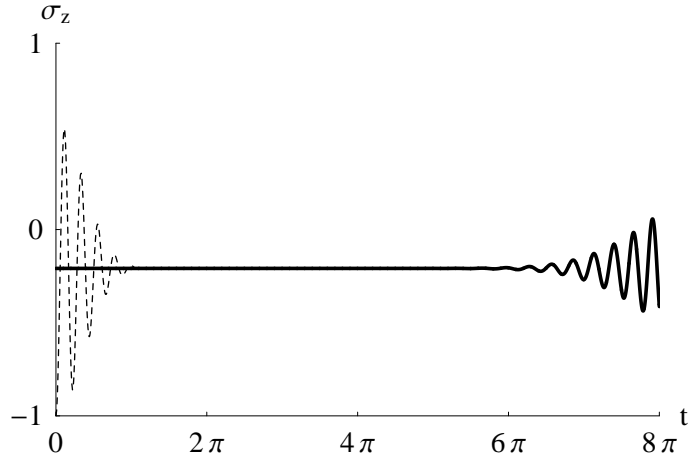}
\end{center}
\caption{The graphs of $\langle\sigma_{z}(t)\rangle$
given by Eqs.~(\ref{probability-ground-state-01})
and (\ref{atomic-population-inversion-equation00})
and $[\mbox{Const.}+4\exp(-\alpha^{2})I^{(0)}_{2}(t)]$ given
by Eqs.~(\ref{definition-Il1t-Il2t-i}) and (\ref{definition-Const})
for $t\in[0,8\pi]$, $\alpha=4$, $\Delta\omega=4$, $\kappa=1$ and $c=4$.
(We consider the off-resonant case.)
A thick solid curve represents $[\mbox{Const.}+4\exp(-\alpha^{2})I^{(0)}_{2}(t)]$
and a thin dashed curve represents $\langle\sigma_{z}(t)\rangle$.
In the numerical calculation of $\langle\sigma_{z}(t)\rangle$
given by Eqs.~(\ref{probability-ground-state-01})
and (\ref{atomic-population-inversion-equation00}),
the summation of the index $n$ is carried out up to $n=100$.
In the numerical calculation of $[\mbox{Const.}+4\exp(-\alpha^{2})I^{(0)}_{2}(t)]$
given by Eqs.~(\ref{definition-Il1t-Il2t-i}) and (\ref{definition-Const}),
the integral $\int_{0}^{\infty}dy$ is replaced with $\int_{0}^{100}dy$.
The interval of the numerical integration $y\in[0,100]$ is divided into $10^{6}$ steps
($\Delta y=1.0\times 10^{-4}$) and we apply Bode's rule.
To obtain the variation of $[\mbox{Const.}+4\exp(-\alpha^{2})I^{(0)}_{2}(t)]$ against the time $t$,
we divide the interval $t\in[0,8\pi]$ into $4000$ steps
($\Delta t=2\pi\times 10^{-3}$)
and we estimate $[\mbox{Const.}+4\exp(-\alpha^{2})I^{(0)}_{2}(t)]$ at each time step.
Looking at this figure, we notice the following facts.
In the graph of $[\mbox{Const.}+4\exp(-\alpha^{2})I^{(0)}_{2}(t)]$, we can observe only the revival
and we cannot observe the initial collapse of the Rabi oscillations.
[$\langle\sigma_{z}(t)\rangle$ shows the initial collapse from $t=0$
until around $t=\pi$.
By contrast, $[\mbox{Const.}+4\exp(-\alpha^{2})I^{(0)}_{2}(t)]$ is nearly equal
to $\mbox{Const.}$ given by Eq.~(\ref{definition-Const}) from $t=0$ to $t=5\pi$.]}
\label{Figure05}
\end{figure}

Second,
we consider the case where the electromagnetic field is off-resonant with the atom
at zero-temperature
for estimating the effects of detuning.
We show graphs of
$[1-2\exp(-\alpha^{2})I^{(0)}_{1}(t)]$
defined in Eq.~(\ref{definition-Il1t-Il2t-i})
and $\langle\sigma_{z}(t)\rangle$ given by Eqs.~(\ref{probability-ground-state-01})
and (\ref{atomic-population-inversion-equation00}) in Fig.~\ref{Figure04}.
We show graphs of
$[\mbox{Const.}+4\exp(-\alpha^{2})I^{(0)}_{2}(t)]$ defined in Eq.~(\ref{definition-Il1t-Il2t-i})
and $\langle\sigma_{z}(t)\rangle$ given by Eqs.~(\ref{probability-ground-state-01})
and (\ref{atomic-population-inversion-equation00}) in Fig.~\ref{Figure05},
where
\begin{equation}
\mbox{Const.}
=
1-\exp(-\alpha^{2})\int_{0}^{\infty}\frac{|\alpha|^{2x}}{\Gamma(1+x)}
(1+\frac{c}{c+x})dx.
\label{definition-Const}
\end{equation}
The above $\mbox{Const.}$ is obtained by replacing
$\cos^{2}(\sqrt{x+c}|\kappa|t)$
and
$\sin^{2}(\sqrt{x+c}|\kappa|t)$
with $1/2$
in the representation of $I^{(0)}_{1}(t)$ given by Eq.~(\ref{definition-Il1t-Il2t-i}).
In these estimations,
we put $\alpha=4$, $\Delta\omega=4$, $\kappa=1$ and $c=4$.
Thus, replacing $\int_{0}^{\infty}dx$ with $\int_{0}^{100}dx$,
we obtain
$\mbox{Const.}=-0.2086...$.

Looking at Figs.~\ref{Figure04} and \ref{Figure05},
we can conclude
that $I^{(0)}_{1}(t)$ only represents the initial collapse
and $I^{(0)}_{2}(t)$ only represents the revival.
The reason why $I^{(0)}_{1}(t)$ given by Eq.~(\ref{definition-Il1t-Il2t-i})
only shows the initial collapse is as follows.
If we let $t$ be a large value ($t\gg 1$),
we can replace the squares of trigonometric functions
[$\cos^{2}(\sqrt{x+c+l}|\kappa|t)$
and
$\sin^{2}(\sqrt{x+c+l}|\kappa|t)$]
included in the integrand of $I^{(0)}_{1}(t)$ with their time average $1/2$.
Thus, we obtain $\mbox{Const.}$ given by Eq.~(\ref{definition-Const}),
and $I^{(0)}_{1}(t)$ never generates the revival of the Rabi oscillations
for $t\gg 1$.

\begin{figure}
\begin{center}
\includegraphics[scale=1.0]{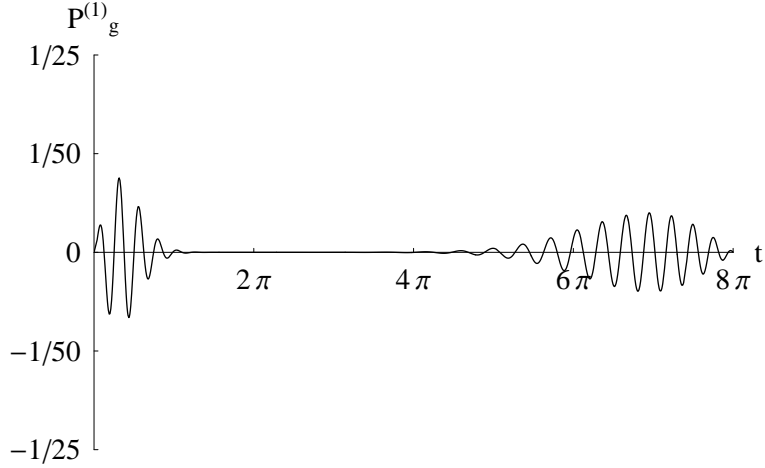}
\end{center}
\caption{The graph of $P^{(1)}_{g}(\beta;t)$ given
by Eqs.~(\ref{P1g-beta-t-final-form}) and (\ref{Qlgt-integral-form})
for $t\in[0,8\pi]$ with putting $\Delta\omega=0$, $\kappa=1$,
$c=0$, $\alpha=4$, $\tilde{\gamma}=1$ and $\theta(\beta)=1/40$.}
\label{Figure06}
\end{figure}

\begin{figure}
\begin{center}
\includegraphics[scale=1.0]{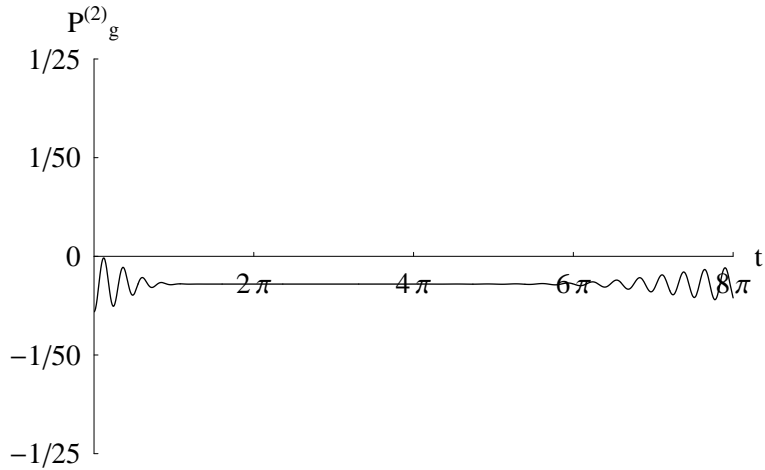}
\end{center}
\caption{The graph of $P^{(2)}_{g}(\beta;t)$ given
by Eqs.~(\ref{Qlgt-integral-form}) and (\ref{P2g-beta-t-final-form})
for $t\in[0,8\pi]$ with putting $\Delta\omega=0$, $\kappa=1$,
$c=0$, $\alpha=4$, $\tilde{\gamma}=1$ and $\theta(\beta)=1/40$.}
\label{Figure07}
\end{figure}

Third, we consider the resonant case at low temperature.
We calculate $P^{(1)}_{g}(\beta;t)$ and $P^{(2)}_{g}(\beta;t)$ with $\Delta\omega=0$,
so that we estimate the effects of low temperature without detuning.
Shown in Eqs.~(\ref{Pgt-integral-form}),
(\ref{P1g-beta-t-final-form}) and (\ref{Qlgt-integral-form}),
we can describe the first-order correction of the atomic population inversion
$P^{(1)}_{g}(\beta;t)$ as a sum of
$I^{(0)}_{1}(t)$, $I^{(0)}_{2}(t)$, $I^{(1)}_{1}(t)$ and $I^{(1)}_{2}(t)$.
Similarly,
from Eqs.~(\ref{Pgt-integral-form}),
(\ref{Qlgt-integral-form}) and (\ref{P2g-beta-t-final-form}),
we can describe the second-order correction
$P^{(2)}_{g}(\beta;t)$ as a sum of
$\{I^{(l)}_{1}(t):l\in\{0,1,2\}\}$
and
$\{I^{(l)}_{2}(t):l\in\{0,1,2\}\}$.
From discussions given in the previous paragraphs,
we can conclude that $I^{(l)}_{1}(t)$ generates the initial collapse
and $I^{(l)}_{2}(t)$ generates the revival for $l=0,1,2,...$.

Putting $\Delta\omega=0$, $\kappa=1$, $c=0$, $\alpha=4$, $\tilde{\gamma}=1$
and $\theta(\beta)=1/40$,
we show graphs of $P^{(1)}_{g}(\beta;t)$ and $P^{(2)}_{g}(\beta;t)$
in Figs.~\ref{Figure06} and \ref{Figure07},
respectively.
Looking at Figs.~\ref{Figure06} and \ref{Figure07},
we find that an amplitude of $P^{(2)}_{g}(\beta;t)$ is comparable
to that of $P^{(1)}_{g}(\beta;t)$.
Thus, the parameters, $\tilde{\gamma}=1$ and $\theta(\beta)=1/40$,
are near the boundary of region where the perturbation theory is effective.

Looking at the graphs of Figs.~\ref{Figure06} and \ref{Figure07}
and the graphs of Figs.~\ref{Figure02} and \ref{Figure03},
we understand that the effect of $P^{(1)}_{g}(\beta;t)$ and $P^{(2)}_{g}(\beta;t)$
on $P_{g}(t)$ is quite small.
Hence, we can conclude that the collapse and the revival of the Rabi oscillations
is robust against the effects of low temperature
in the region where the second-order perturbation theory is effective.

\section{\label{section-Discussions}Discussions}

In this paper, we separate the atomic population inversion of the Jaynes-Cummings model
into two integrals using the Abel-Plana formula.
By numerical calculations, we show that the first integral represents
the initial collapse (the semi-classical limit)
and the second integral represents the revival (the quantum correction).
Moreover, we examine the time-evolution of the JCM
with the initial thermal coherent state for the cavity mode at low temperature
by the second-order perturbation theory.
We describe the first and second-order corrections as sums of integrals,
using the Abel-Plana formula.

The Abel-Plana formula and its generalized versions are often made use of
for the calculations of the Casimir energies
in different configurations~\cite{Bordag2001,Inui2003,Saharian2000-Preprint}.
The author thinks that the Abel-Plana formula
has a wide application in the field of the quantum optics.

\section*{Acknowledgments}

The author thanks colleagues of IMS Lab. Inc. for encouragement.

\appendix

\section{\label{appendix-Abel-Plana-formula}
The derivation of the Abel-Plana formula}

In this section, we give details of the derivation of the summation formula
described in Eq.~(\ref{Abel-Plana-formula-0}).
(This formula is introduced in Ref.~\cite{Whittaker1927}
without precise derivation.
The author of Ref.~\cite{Whittaker1927} only mentions a suggestion
for proving it.)

\begin{figure}
\begin{center}
\includegraphics[scale=1.0]{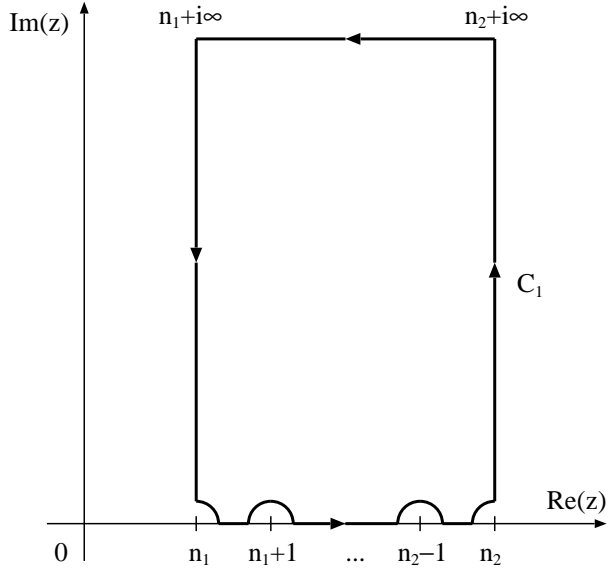}
\end{center}
\caption{The closed contour $C_{1}$ defined on the complex plane.}
\label{Figure08}
\end{figure}

At first, we assume $\phi(z)$ to be analytical and bounded for all complex values of $z$
such that $n_{1}\leq\mbox{Re}(z)\leq n_{2}$,
where $n_{1}$ and $n_{2}$ are certain integers.
Then, $\phi(z)/(e^{-2\pi iz}-1)$ is analytical and bounded $\forall z$
such that $n_{1}\leq\mbox{Re}(z)\leq n_{2}$
except for $z=n_{1},n_{1}+1,n_{1}+2,...,n_{2}-1,n_{2}$.
Hence, if we think about the closed contour $C_{1}$ shown in Fig.~\ref{Figure08},
$\phi(z)/(e^{-2\pi iz}-1)$ is analytical and bounded inside $C_{1}$,
so that
\begin{equation}
\oint_{C_{1}}\frac{\phi(z)}{e^{-2\pi iz}-1}dz=0.
\label{Cauchy-integral-C1}
\end{equation}
(This is the Cauchy integral theorem.)

\begin{figure}
\begin{center}
\includegraphics[scale=1.0]{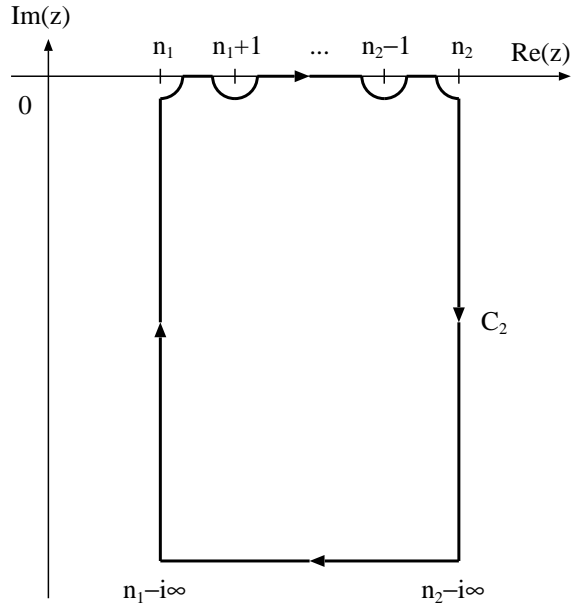}
\end{center}
\caption{The closed contour $C_{2}$ defined on the complex plane.}
\label{Figure09}
\end{figure}

Similarly, $\phi(z)/(e^{2\pi iz}-1)$ is analytical and bounded $\forall z$
such that $n_{1}\leq\mbox{Re}(z)\leq n_{2}$
except for $z=n_{1},n_{1}+1,n_{1}+2,...,n_{2}-1,n_{2}$.
Hence, if we think about the closed contour $C_{2}$ shown in Fig.~\ref{Figure09},
$\phi(z)/(e^{2\pi iz}-1)$ is analytical and bounded inside $C_{2}$,
so that
\begin{equation}
\oint_{C_{2}}\frac{\phi(z)}{e^{2\pi iz}-1}dz=0.
\label{Cauchy-integral-C2}
\end{equation}

\begin{figure}
\begin{center}
\includegraphics[scale=1.0]{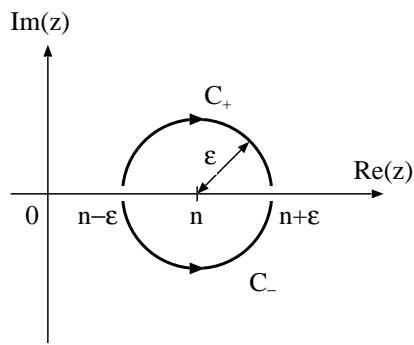}
\end{center}
\caption{The paths $C_{+}$ and $C_{-}$ defined on the complex plane.
Describing $\bar{C}_{+}$ as an opposite path of $C_{+}$,
$\bar{C}_{+}$ and $C_{-}$ form a closed contour.}
\label{Figure10}
\end{figure}

Next, we consider the following integrals:
\begin{equation}
I(n)
=\int_{C_{+}}\frac{\phi(z)}{e^{-2\pi iz}-1}dz
+\int_{C_{-}}\frac{\phi(z)}{e^{2\pi iz}-1}dz,
\label{derivation-Abel-Plana-integral-residue-01}
\end{equation}
where $C_{+}$ and $C_{-}$ are paths on the complex plane shown in Fig.~\ref{Figure10}.
In Fig.~\ref{Figure10}, $n$ represents an integer such that
$n_{1}<n<n_{2}$ and $\epsilon$ is a small positive infinitesimal quantity.
We can rewrite $I(n)$ defined in Eq.~(\ref{derivation-Abel-Plana-integral-residue-01}) as
\begin{eqnarray}
I(n)
&=&
-\int_{\bar{C}_{+}}\frac{\phi(z)}{e^{-2\pi iz}-1}dz
+\int_{C_{-}}\frac{\phi(z)}{e^{2\pi iz}-1}dz \nonumber \\
&=&
\int_{\bar{C}_{+}}
\frac{e^{2\pi iz}\phi(z)}{e^{2\pi iz}-1}dz
+\int_{C_{-}}\frac{\phi(z)}{e^{2\pi iz}-1}dz \nonumber \\
&=&
\oint_{\bar{C}_{+}+C_{-}}
\frac{\Phi(z)}{e^{2\pi iz}-1}dz,
\label{derivation-Abel-Plana-integral-residue-02}
\end{eqnarray}
where $\bar{C}_{+}$ is an opposite path of $C_{+}$ and $\Phi(z)$ is given by
\begin{equation}
\Phi(z)
=
\left\{
\begin{array}{ll}
e^{2\pi iz}\phi(z) & \mbox{for $\mbox{Im}(z)\geq 0$} \\
\phi(z) & \mbox{for $\mbox{Im}(z)<0$}
\end{array}
\right..
\end{equation}
Inside of the closed contour $\bar{C}_{+}+C_{-}$,
$\Phi(z)/(e^{2\pi iz}-1)$ has only one pole, $z=n$.

If we write $z=n+u$, we can expand
$\Phi(z)/(e^{2\pi iz}-1)$ around $z=n$ as
\begin{equation}
\frac{\Phi(n+u)}{e^{2\pi i(n+u)}-1}
=
\frac{\phi(n)}{2\pi i}\frac{1}{u}
+
P(u),
\end{equation}
where $P(u)$ is a power series that includes only terms of nonnegative degrees,
so that the residue is given by
\begin{equation}
\mbox{Res}[\Phi(z)/(e^{2\pi iz}-1),n]=\phi(n),
\end{equation}
and we obtain
\begin{equation}
I(n)=\phi(n).
\label{derivation-Abel-Plana-integral-residue-03}
\end{equation}

\begin{figure}
\begin{center}
\includegraphics[scale=1.0]{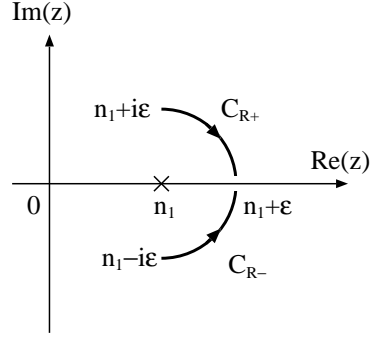}
\end{center}
\caption{The paths $C_{\mbox{\scriptsize R}+}$ and $C_{\mbox{\scriptsize R}-}$
defined on the complex plane.}
\label{Figure11}
\end{figure}

Furthermore, we think about the following integrals:
\begin{equation}
I_{\mbox{\scriptsize R}}(n_{1})
=\int_{C_{\mbox{\scriptsize R}+}}\frac{\phi(z)}{e^{-2\pi iz}-1}dz
+\int_{C_{\mbox{\scriptsize R}-}}\frac{\phi(z)}{e^{2\pi iz}-1}dz,
\label{derivation-Abel-Plana-integral-residue-R1}
\end{equation}
where $C_{\mbox{\scriptsize R}+}$ and $C_{\mbox{\scriptsize R}-}$ are
paths on the complex plane shown in Fig.~\ref{Figure11}.
In Fig.~\ref{Figure11}, $\epsilon$ is a small positive infinitesimal quantity.
[The index R in $I_{\mbox{\scriptsize R}}(n_{1})$, $C_{\mbox{\scriptsize R}+}$
and $C_{\mbox{\scriptsize R}-}$ implies that
$C_{\mbox{\scriptsize R}+}$ and $C_{\mbox{\scriptsize R}-}$
are right halves of $C_{+}$ and $C_{-}$, respectively.]
In the limit of $\epsilon\to +0$,
we obtain the following relation:
\begin{equation}
\lim_{\epsilon\to +0}
I_{\mbox{\scriptsize R}}(n_{1})
=
\frac{1}{2}
\lim_{\epsilon\to +0}
I(n_{1})
=
\frac{1}{2}\phi(n_{1}).
\label{derivation-Abel-Plana-integral-residue-R2}
\end{equation}

\begin{figure}
\begin{center}
\includegraphics[scale=1.0]{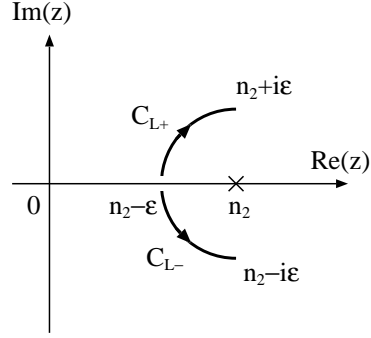}
\end{center}
\caption{The paths $C_{\mbox{\scriptsize L}+}$ and $C_{\mbox{\scriptsize L}-}$
defined on the complex plane.}
\label{Figure12}
\end{figure}

In a similar way,
we obtain the following relation:
\begin{equation}
I_{\mbox{\scriptsize L}}(n_{1})
=\int_{C_{\mbox{\scriptsize L}+}}\frac{\phi(z)}{e^{-2\pi iz}-1}dz
+\int_{C_{\mbox{\scriptsize L}-}}\frac{\phi(z)}{e^{2\pi iz}-1}dz,
\label{derivation-Abel-Plana-integral-residue-L1}
\end{equation}
where $C_{\mbox{\scriptsize L}+}$ and $C_{\mbox{\scriptsize L}-}$ are paths
on the complex plane shown in Fig.~\ref{Figure12}
and
\begin{equation}
\lim_{\epsilon\to +0}
I_{\mbox{\scriptsize L}}(n_{2})
=
\frac{1}{2}
\lim_{\epsilon\to +0}
I(n_{2})
=
\frac{1}{2}\phi(n_{2}).
\label{derivation-Abel-Plana-integral-residue-L2}
\end{equation}

From Eqs.~(\ref{Cauchy-integral-C1}), (\ref{Cauchy-integral-C2}),
(\ref{derivation-Abel-Plana-integral-residue-03}),
(\ref{derivation-Abel-Plana-integral-residue-R2})
and (\ref{derivation-Abel-Plana-integral-residue-L2}),
we obtain
\begin{eqnarray}
&&
\frac{1}{2}\phi(n_{1})
+\sum_{n=n_{1}+1}^{n_{2}-1}\phi(n)
+\frac{1}{2}\phi(n_{2}) \nonumber  \\
&&
+
(\int_{n_{1}+\epsilon}^{n_{1}+1-\epsilon}
+\int_{n_{1}+1+\epsilon}^{n_{1}+2-\epsilon}
+...
+\int_{n_{2}-1+\epsilon}^{n_{2}-\epsilon})
[
\frac{\phi(x)}{e^{-2\pi ix}-1}
+
\frac{\phi(x)}{e^{2\pi ix}-1}
]dx \nonumber \\
&&
+i
\int_{\epsilon}^{\infty}\frac{\phi(n_{2}+iy)}{e^{2\pi y}-1}dy
-
\int_{n_{1}}^{n_{2}}\frac{\phi(x+i\infty)}{e^{-2\pi i(x+i\infty)}-1}dx \nonumber \\
&&
-i
\int_{\epsilon}^{\infty}\frac{\phi(n_{1}+iy)}{e^{2\pi y}-1}dy
-i
\int_{\epsilon}^{\infty}\frac{\phi(n_{2}-iy)}{e^{2\pi y}-1}dy \nonumber \\
&&
-
\int_{n_{1}}^{n_{2}}\frac{\phi(x-i\infty)}{e^{2\pi i(x-i\infty)}-1}dx
+i
\int_{\epsilon}^{\infty}\frac{\phi(n_{1}-iy)}{e^{2\pi iy}-1}dy
=0.
\label{Abel-Plana-formula-Whittaker-Watson-00}
\end{eqnarray}

Looking at Eq.~(\ref{Abel-Plana-formula-Whittaker-Watson-00}),
we notice the following facts.
Because $\phi(z)$ is bounded for all $z$ such that $n_{1}\leq\mbox{Re}(z)\leq n_{2}$,
we obtain
\begin{eqnarray}
\int_{n_{1}}^{n_{2}}
\frac{\phi(x+i\infty)}{e^{-2\pi i(x+i\infty)}-1}dx
&=&
0, \nonumber \\
\int_{n_{1}}^{n_{2}}
\frac{\phi(x-i\infty)}{e^{2\pi i(x-i\infty)}-1}dx
&=&
0.
\end{eqnarray}
The integrand of the fourth term in the left-hand side of
Eq.~(\ref{Abel-Plana-formula-Whittaker-Watson-00})
can be rewritten as
\begin{eqnarray}
\frac{\phi(x)}{e^{-2\pi ix}-1}+\frac{\phi(x)}{e^{2\pi ix}-1}
&=&
\frac{(e^{2\pi ix}-1)\phi(x)+(e^{-2\pi ix}-1)\phi(x)}
{(e^{-2\pi ix}-1)(e^{2\pi ix}-1)} \nonumber \\
&=&
\frac{(e^{2\pi ix}+e^{-2\pi ix}-2)\phi(x)}{1-e^{-2\pi ix}-e^{2\pi ix}+1} \nonumber \\
&=&
-\phi(x),
\end{eqnarray}
so that the limit of the integral,
as $\epsilon$ approaches zero ($\epsilon\to +0$),
is equal to
\begin{eqnarray}
&&
\lim_{\epsilon\to +0}
(\int_{n_{1}+\epsilon}^{n_{1}+1-\epsilon}
+\int_{n_{1}+1+\epsilon}^{n_{1}+2-\epsilon}
+...
+\int_{n_{2}-1+\epsilon}^{n_{2}-\epsilon})
[
\frac{\phi(x)}{e^{-2\pi ix}-1}
+
\frac{\phi(x)}{e^{2\pi ix}-1}
]dx \nonumber \\
&=&
-\int_{n_{1}}^{n_{2}}\phi(x)dx.
\end{eqnarray}

Hence, we can rewrite Eq.~(\ref{Abel-Plana-formula-Whittaker-Watson-00}) as
\begin{eqnarray}
&&
\frac{1}{2}\phi(n_{1})
+\sum_{n=n_{1}+1}^{n_{2}-1}\phi(n)
+\frac{1}{2}\phi(n_{2}) \nonumber  \\
&=&
\int_{n_{1}}^{n_{2}}\phi(x)dx \nonumber  \\
&&
+i
\int_{0}^{\infty}
\frac{
-\phi(n_{2}+iy)+\phi(n_{1}+iy)+\phi(n_{2}-iy)-\phi(n_{1}-iy)
}{e^{2\pi y}-1}dy.
\end{eqnarray}
Thus, we obtain the formula of Eq.~(\ref{Abel-Plana-formula-0}).
If $\phi(z)\to 0$ as $\mbox{Re}(z)\to +\infty$,
we obtain
\begin{equation}
\frac{1}{2}\phi(0)
+\sum_{n=1}^{\infty}\phi(n)
=
\int_{0}^{\infty}\phi(x)dx
+i
\int_{0}^{\infty}
\frac{
\phi(iy)-\phi(-iy)
}{e^{2\pi y}-1}dy.
\end{equation}
So that, we obtain the formula of Eq.~(\ref{Abel-Plana-formula-1}).
This equation is called the Abel-Plana formula.

\section{\label{appendix-remarks-numerical-calculations}
Some remarks about numerical calculations}

In section~\ref{section-Properties-integrals},
for the numerical calculations of $J_{1}(t)$, $J_{2}(t)$,
$I^{(l)}_{1}(t)$ and $I^{(l)}_{2}(t)$,
we use the Fortran compiler with quadruple-precision complex
(a pair of quadruple-precision real numbers).

We evaluate the Gamma function included
in Eqs.~(\ref{definition-Il1t-Il2t-i}) and (\ref{definition-first-and-second-integral00})
numerically
by the Lanczos approximation~\cite{Press1992},
\begin{eqnarray}
\ln\Gamma(z)
&=&
(z+\frac{1}{2})\ln(z+g+\frac{1}{2})-(z+g+\frac{1}{2}) \nonumber \\
&&
+\ln[\frac{\sqrt{2\pi}}{z}(c_{0}+\sum_{n=1}^{6}\frac{c_{n}}{z+n}+\epsilon)]
\quad\quad
\mbox{for $\mbox{Re}(z)>0$},
\label{Lanczos-formula}
\end{eqnarray}
where $g=5$,
\begin{eqnarray}
c_{0}
&=&
1.000\mbox{ }000\mbox{ }000\mbox{ }190\mbox{ }015, \nonumber \\
c_{1}
&=&
76.180\mbox{ }091\mbox{ }729\mbox{ }471\mbox{ }46, \nonumber \\
c_{2}
&=&
-86.505\mbox{ }320\mbox{ }329\mbox{ }416\mbox{ }77, \nonumber \\
c_{3}
&=&
24.014\mbox{ }098\mbox{ }240\mbox{ }830\mbox{ }91, \nonumber \\
c_{4}
&=&
-1.231\mbox{ }739\mbox{ }572\mbox{ }450\mbox{ }155, \nonumber \\
c_{5}
&=&
0.120\mbox{ }865\mbox{ }097\mbox{ }386\mbox{ }617\mbox{ }9\times 10^{-2} \nonumber \\
c_{6}
&=&
-0.539\mbox{ }523\mbox{ }938\mbox{ }495\mbox{ }3\times 10^{-5}.
\end{eqnarray}
The approximation with Eq.~(\ref{Lanczos-formula}) gives the error upper bound
$|\epsilon|<2\times 10^{-10}$.

For carrying out the numerical integration of
$J_{1}(t)$ defined in Eq.~(\ref{definition-first-and-second-integral00})
and $I^{(l)}_{1}(t)$ defined in Eq.~(\ref{definition-Il1t-Il2t-i}),
we use Simpson's rule~\cite{Press1992}.
For carrying out the numerical integration of
$J_{2}(t)$ defined in Eq.~(\ref{definition-first-and-second-integral00})
and $I^{(l)}_{2}(t)$ defined in Eq.~(\ref{definition-Il1t-Il2t-i}),
we use Bode's rule~\cite{Press1992}.

In the numerical integration of $J_{2}(t)$,
we pay attention to the following fact.
In the limit as $y\to +0$,
the integrand of $J_{2}(t)$ converges on a finite value as
\begin{equation}
\lim_{y\to +0}
\frac{1}{e^{2\pi y}-1}
\mbox{Im}\{
\frac{|\alpha|^{2iy}}{\Gamma(1+iy)}\cos(2\sqrt{iy}t)
\}
=
\frac{1}{2\pi}(2\ln|\alpha|+\gamma-2t^{2}),
\end{equation}
where $\gamma=0.577215...$ is the Euler-Mascheroni constant.
Similarly, in the limit as $y\to +0$,
the integrand of $I^{(0)}_{2}(t)$ converges on a finite value as
\begin{eqnarray}
&&
\lim_{y\to +0}
\frac{1}{e^{2\pi y}-1}
\mbox{Im}\{
\frac{|\alpha|^{2iy}}{\Gamma(1+iy)}
[\cos^{2}(\sqrt{c+iy}|\kappa|t)
+
\frac{c}{c+iy}\sin^{2}(\sqrt{c+iy}|\kappa|t)]
\} \nonumber \\
&=&
\frac{1}{4c\pi}[-1+2c\gamma+\cos(2\sqrt{c}|\kappa|t)+4c\ln|\alpha|].
\end{eqnarray}
As shown above,
when we calculate $I^{(l)}_{2}(t)$ numerically,
we have to be careful about taking the limit of the integrand of $I^{(l)}_{2}(t)$
as $y\to +0$.

The numerical integration of $J_{2}(t)$ is more difficult
than that of $J_{1}(t)$.
The numerical evaluation of $J_{2}(t)$ for $t>8\pi$
does not converge on a reasonable value
even if we use Romberg's method~\cite{Press1992}.
The exactly same things happen
when we calculate $I^{(0)}_{1}(t)$ and $I^{(0)}_{2}(t)$ numerically.

\end{document}